\documentclass[preprint,showpacs,aps]{revtex4}

\usepackage{graphicx}% Include figure files
\usepackage{epsfig}
\usepackage{epstopdf}
\usepackage{dcolumn}% Align table columns on decimal point
\usepackage{bm}% bold math
\usepackage{amsmath}% AMS math commands
\usepackage{amssymb}
\usepackage{placeins} %to use \FloatBarrier command

\def\Journal#1#2#3#4{{#1} {\bf#2}, {#3} {(#4)}}
\def\NP{{ Nucl. Phys.} }
\def\PLB{{ Phys. Lett.}  B}
\def\PL{{ Phys. Lett.}}

\def\PPNP{{ Prog. Part. Nucl. Phys.}}
\def\PRP{{ Phys. Rep.}}
\def\PRL{ Phys. Rev. Lett.}
\def\PR{{ Phys. Rev.}}
\def\PRD{{ Phys. Rev.} D}
\def\PRC{{ Phys. Rev.} C}

\def\ZPC{{Z. Phys.} C}

\def\EPJA{{Eur. Phys. J.} A}
\def\EPJC{{Eur. Phys. J.} C}

\def\ZPC{{Z. Phys.} C}

\def\MPLA{{Mod. Phys. Lett.} A}
\def\CPC{Comput. Phys. Commun.}
\def\IJMP{Int. J. Mod. Phys.}

\def\ra{\rightarrow}

\def\be{\begin{equation}}
\def\ee{\end{equation}}
\def\bea{\begin{eqnarray}}
\def\eea{\end{eqnarray}}

\def\ua{\uparrow}
\def\da{\downarrow}

\def\ubar{{\bar u}}
\def\dbar{{\bar d}}
\def\sbar{{\bar s}}

\def\NP{{ Nucl. Phys.}}
\def\ANP{{Adv. Nucl. Phys.}}
\def\APP{{ Acta Phys. Pol.}}

\begin{document}
\title{The spin dependent structure functions of the nucleon}

\author{F. Bissey}
\email{f.r.bissey@massey.ac.nz}
\author{Fu-Guang Cao}
\email{f.g.cao@massey.ac.nz}
\author{A. I. Signal}
\email{a.i.signal@massey.ac.nz}
\affiliation{Institute of Fundamental Sciences PN461 \\ Massey University \\
Private Bag 11 222,  Palmerston North \\
New Zealand}

\vskip 0.5cm
\begin{abstract}
We calculate the spin dependent structure functions $g_{1}(x)$ and $g_{2}(x)$ 
of the proton and neutron.
Our calculation uses the meson cloud model of nucleon structure, which has previously 
given a good description of the HERMES data on polarized sea quark distributions, and 
includes all the leading contributions to spin dependent effects in this model. 
We find good agreement between our calculations and the current experimental data for the 
structure functions.
We include in our calculations kinematic terms, which mix transverse and longitudinal 
spin components, for hadrons of spin 1/2, 1 and 3/2, and which can give considerable 
contributions to the $g_{2}$ structure functions.
We also consider the possible interference terms between baryons or mesons in different 
final states with the same quantum numbers, and show that most of these terms do not give 
leading contributions to the spin dependent structure functions.

\end{abstract}


\maketitle

\setcounter{footnote}{0}
\section{Introduction}

The spin dependent structure functions of the nucleon are the subject of much 
theoretical and experimental interest. 
The main reason for this interest has been the large amount of evidence, starting 
with the EMC experiment \cite{EMC} which strongly suggests that constituent quark 
models cannot fully describe the spin structure of protons and neutrons. 
This has lead to considerable activity in order to determine how the spin of 
nucleons is built up from the intrinsic spin and orbital angular momentum of 
their constituent quarks and gluons.  

Since 1988 further polarised deep inelastic scattering experiments have generally 
confirmed the EMC results for the proton - photon asymmetry $A_{1}$ and proton spin 
structure function $g_{1}(x)$ \cite{g1pE154, g1pSMC, g1p_HERMES}.
These measurements have also been performed on deuteron and neutron 
targets \cite{A1g1_SMC, g1n_HERMES}, which has enabled the Bjorken sum rule \cite{Bj} to be 
tested at the five per cent level. 

In addition, there have been measurements, using transversely polarised targets, of 
the second nucleon - photon asymmetries $A^{N}_{2}$ and the related structure functions 
$g^{N}_{2}(x)$ \cite{E143, A2g2_E155, A2E154, g2HallA, g2HallA_Kramer}. 
The $g_{2}(x)$ structure functions are of interest because they do not have a simple 
interpretation in the quark-parton model, but are related to transverse momentum of quarks 
and higher twist operators which measure correlations between quarks and gluons. 
The identification of a higher twist component in a measurement of $g_{2}^{N}(x)$ would 
be significant as this would give new information on the gluon field inside the nucleon, 
and its relationship with the quark fields.  

Recently new experimental approaches have sought to augment the information available 
from deep inelastic scattering (DIS) experiments. 
These include semi-inclusive polarised DIS (HERMES and COMPASS), polarised 
proton-proton collisions (RHIC) and polarised photoproduction.

There have been a number of theoretical approaches to calculating $g_{1}(x)$ and 
$g_{2}(x)$ using phenomenological models of nucleon structure such as the 
MIT bag model \cite{SST, Song, Stratmann} and the chiral soliton model \cite{XSM, Bochum, JapXSM}.
In addition there have been lattice calculations of some of the nucleon matrix elements of 
operators corresponding to small moments of the structure functions \cite{Negele}.

In this paper we shall use the meson cloud model (MCM) to calculate the spin dependent 
structure functions of the nucleon.
This model has been applied successfully in spin independent DIS, giving a good description 
of the HERA data on semi-inclusive DIS with a leading neutron \cite{H1n, Zeusn}, and 
also dijet events with a leading neutron \cite{Zeus2j, H12j}. 
In addition the MCM gives a good description of the observed violation \cite{NMC, E866} 
of the Gottfried sum rule \cite{Kum98, GP, WMelnitchoukST}.

The meson cloud model (MCM) has been used previously to calculate 
$g_{1}(x)$ \cite{HHoltmannSS, CBorosT}.
In those calculations pseudoscalar mesons were identified as the main constituents of the 
meson cloud. 
While these mesons do not directly contribute to the structure function, the presence of the 
cloud transfers some angular momentum from the quarks in the `bare' proton to the meson 
cloud and results in a decrease in the calculated first moment of $g_{1}^{p}$ compared to the 
MIT bag model.
More recently it has been realised that vector mesons, particularly the $\rho$, can 
also play a role in the spin structure of the proton \cite{FS}. 
In particular these will give rise to flavour symmetry breaking in the sea, and the 
$\Delta \bar{u}(x) - \Delta \bar{d}(x)$ difference has been calculated by a number of 
authors \cite{FCaoS_plzdsea, KM}.
Interestingly these calculations predict that the spin dependent symmetry breaking is 
quite small, in contrast to the spin independent symmetry breaking combination 
$\bar{u}(x) - \bar{d}(x)$ which is observed to be large. 
Recently these calculations were extended to the spin dependent sea distributions 
($\Delta \ubar,\Delta \dbar,\Delta s, \Delta \sbar$) \cite{FCaoS_ps2}, and were 
found to be in good agreement with the recent results from HERMES \cite{HERMES02}.

In this paper we revisit the earlier calculations of $g_{1}^{N}(x)$ in light of the 
developments in the MCM since that time. 
We also extend the calculations of structure functions using the MCM to $g_{2}^{N}(x)$, 
and investigate the kinematic regions where it may be possible to observe a twist-3 
piece of the structure functions.

In Section 2 of this paper we present the formalism for discussing spin dependent 
structure functions in the meson cloud model, including a discussion of the kinematic terms 
which lead to $g_{1}$ of cloud components contributing to $g_{2}$ of the nucleon (and vice 
versa) \cite{KM}.
Contributions to the structure functions from interference between different states of the 
cloud \cite{FCaoS_plzdsea} are discussed in Section 3, and it is shown that most of the leading 
interference contributions vanish. 
In Section 4 we apply the MCM formalism and determine all the necessary momentum 
distributions of the components of the meson cloud. 
We also discuss the correct prescription to use in describing the energy of the intermediate 
state hadrons in the MCM.
In Section 5 we calculate the spin dependent structure functions of the baryons and mesons 
in the cloud using the MIT bag model.
The numerical results for the nucleon structure functions are shown and discussed in 
Section 6.
In the last section we summarise our findings.

\section{Spin dependent structure functions in the meson cloud model}

In the LAB frame the cross section for inclusive inelastic lepton-nucleon scattering 
may be written in terms of the product of lepton and hadron tensors
\bea
\frac{d^{2}\sigma}{d\Omega dE'} = \frac{\alpha^{2}}{q^{4}}\frac{E'}{E}
L_{\mu \nu} W^{\mu \nu}
\eea 
where $\alpha$ is the fine structure constant, $E (E')$ is the energy of the 
incident (scattered) lepton and $q^{2}$ is the squared four-momentum transfer.
In spin dependent (polarised) scattering we are interested in the antisymmetric 
part of the hadron tensor $W_{\mu \nu}^{A}$, which can be written in terms of two 
structure functions $G_{1}$ and $G_{2}$ \cite{Roberts}
\bea
W_{\mu \nu}^{A} = \frac{i}{m_{N}} \epsilon_{\mu \nu \rho \sigma} q^{\rho} 
\left[ s_{N}^{\sigma} G_{1}(\nu, Q^{2}) + 
\frac{s_{N}^{\sigma} p_{N} \cdot q - p_{N}^{\sigma} s_{N} \cdot q}{m_{N}^{2}} G_{2}(\nu, Q^{2}) \right].
\label{eq:WA1}
\eea
Here $\nu$ is the energy transfer, $Q^{2} = -q^{2}$, and $s_{N}^{\mu}$ is the nucleon spin 
vector, normalised to $s_{N}^{2} = -m_{N}^{2}$.
In the Bjorken limit ($Q^{2}, \nu \ra \infty$) the structure functions scale, 
modulo perturbative QCD logarithmic evolution in $Q^{2}$,
\bea
\frac{\nu}{m_{N}} G_{1}(\nu, Q^{2}) & \ra & g_{1}(x)  \nonumber \\
\frac{\nu^{2}}{m_{N}^{2}} G_{2}(\nu, Q^{2}) & \ra & g_{2}(x) 
\eea
where the scaling variable $x = Q^{2}/(2 m_{N} \nu)$ lies between 0 and 1.
In this limit we have 
\bea
W_{\mu \nu}^{A} = i \epsilon_{\mu \nu \rho \sigma} q^{\rho} 
\left[ \frac{s_{N}^{\sigma}}{p_{N} \cdot q} g_{1}(x) + 
\frac{s_{N}^{\sigma} p_{N} \cdot q - p_{N}^{\sigma} s_{N} \cdot q}{(p_{N} \cdot q)^{2}} g_{2}(x) \right].
\label{eq:WA2}
\eea

In order to discuss the structure functions separately we use a projection operator
\bea
P_{\mu \nu} = \frac{i}{2 p_{N} \cdot q} \epsilon_{\mu \nu \alpha \beta} 
q^{\alpha}s_{N}^{\beta},
\label{eq:proj}
\eea 
such that 
\bea
P^{\mu \nu} W_{\mu \nu}^{A}(p_{N}, s_{N}, q) = 
\frac{m_{N}^{2} q^{2} + (s_{N} \cdot q)^{2}}{(p_{N} \cdot q)^{2}} g_{1}(x) - \gamma^{2} g_{2}(x)
\eea
where
\bea
\gamma^{2} = \frac{4 x^{2} m_{N}^{2}}{Q^{2}}.
\eea

\begin{figure}[htb]
\begin{center}
\includegraphics[width=10 cm]{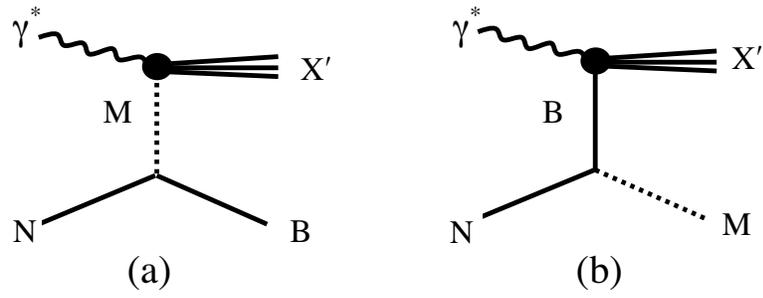}
%\vskip -0.5cm
\caption{The photon may be scattered from (a) virtual meson and (b) virtual baryon.}
\label{fig:MCM}
\end{center}
\end{figure}

In the meson cloud model (MCM) \cite{Sullivan, AThomas83} the nucleon can be 
viewed as a bare nucleon plus some baryon-meson Fock states which result from
the fluctuation of nucleon to baryon plus meson $N \ra M B$.
The wavefunction of the nucleon can be written as \cite{HHoltmannSS},
\begin{equation}
\begin{split}
|N\rangle_{\rm physical} &=  \sqrt{Z} |N\rangle_{\rm bare} \\
& \quad + \sum_{MB} \sum_{\lambda_{M} \lambda_{B}} 
\int dy \, d^2 {\bf k}_\perp \, \phi^{\lambda_{M} \lambda_{B}}_{MB}(y,k_\perp^2)
\, |M^{\lambda_{M}}(y, {\bf k}_\perp); B^{\lambda_{B}}(1-y,-{\bf k}_\perp)
\rangle. 
\label{eq:NMCM}
\end{split}
\end{equation}
Here $Z$ is the wave function renormalization constant and
$\phi^{\lambda_{M} \lambda_{B}}_{MB}(y,k_\perp^2)$ 
is the wave function of the Fock state containing a meson ($M$)
with longitudinal momentum fraction $y$, transverse momentum ${\bf k}_\perp$,
and helicity $\lambda_{M}$, and a baryon ($B$) with momentum fraction $1-y$,
transverse momentum $-{\bf k}_\perp$, and helicity $\lambda_{B}$.
The model assumes that the lifetime of a virtual baryon-meson Fock state is much
longer than the interaction time in the deep inelastic or Drell-Yan
process, thus scattering from the virtual baryon-meson Fock states
can contribute to the observed structure functions of the nucleon, as shown in 
figure~\ref{fig:MCM}.

The contribution to the nucleon tensor $W_{\mu \nu}$ from processes such as that in  
figure~\ref{fig:MCM}, where the virtual photon interacts with a component of the cloud 
(such as a $\rho$ meson) is given by 
\bea
\delta W_{\mu \nu} = \int \frac{d^{3}p_{B}}{(2\pi)^{3}} \frac{m_{M}}{m_{B}} 
\sum_{\lambda,\lambda^\prime} |J_{NMB}|^{2} W_{\mu \nu}^{M}(k, s_{M}, q)
\label{eq:deltaW}
\eea
where $s_{M}$ is the meson spin vector (normalised to $s_{M}^{2} = -m_{M}^{2}$), 
and $J_{NMB}(p_{N}, k, p_{B}, s_{N}, s_{M}, s_{B})$ 
is the meson propagator multiplied by the $NMB$ vertex.
The meson tensor here is defined by
\bea
W_{\mu \nu}^{M}(k, s_{M}, q) = \frac{1}{2\pi} \sum_{X}(2\pi)^{4} \delta^{4} (k + q - p_{X}) 
\langle k, s_{M}|J_{\mu}|X \rangle \langle X|J_{\nu}|k, s_{M} \rangle.
\eea
As first shown by Sullivan \cite{Sullivan}, the meson contribution to the nucleon tensor 
is expressed in terms of the meson tensor and the nucleon-meson-baryon vertex, and this leads 
to the contribution being expressed as a convolution between the meson tensor and the 
probability distribution for finding the meson in the cloud with momentum fraction $y$, eg
\bea
\delta F_{2}^{N}(x) = \int_x^1 dy f_{\rho N/N}(y) F_{2}^{\rho}\left( \frac{x}{y} \right)
\eea
gives the contribution to the nucleon structure function $F_{2}$ arising from the virtual photon 
interacting with the $\rho$ meson from the $N \ra N \rho$
part of the meson cloud.

Baryon or meson components of the cloud with spin $\geq 1/2$ will contribute directly to the 
antisymmetric part of the nucleon tensor.
We consider three cases of interest.

\subsection{Spin $1/2$ Baryons}

A spin $1/2$ baryon in the cloud, such as a nucleon or a $\Lambda$, has the antisymmetric 
part of its tensor similar to equation (\ref{eq:WA2}), with the nucleon mass, momentum and 
spin vector replaced by $m_{B}$, $k$ (the baryon four momentum), and $s_{B}$ respectively.
Multiplying by the projection operator $\tilde{P}^{\mu \nu} = (m_{N}/m_{B}) P^{\mu \nu}$, 
where $P^{\mu \nu}$ is given in  equation (\ref{eq:proj}), gives
\bea
\tilde{P}^{\mu \nu} W_{\mu \nu}^{(1/2) A}(k, s_{B}, q) = 
\frac{m_{N}}{m_{B}} [A_{1}g_{1}^{B}(k,q) + A_{2}g_{2}^{B}(k,q)]
\eea
where we have the coefficients
\bea
A_{1} & = & \frac{1}{p_{N}\cdot q \, k\cdot q} 
(s_{N}\cdot q \, s_{B}\cdot q - q^{2} s_{N}\cdot s_{B}),  
\label{eq:Acoeffs1} \\
A_{2} & = & \frac{q^{2}}{p_{N}\cdot q \, (k\cdot q)^{2}} 
(s_{N}\cdot k \, s_{B}\cdot q - k \cdot q \, s_{N}\cdot s_{B}) 
\label{eq:Acoeffs2}
\eea
and the structure functions $g_{i}^{B}$ are those for the spin $1/2$ baryon. 
In what follows we will use time ordered perturbation theory, which has the advantage 
that all the baryon and meson structure functions that are required are those 
for on-shell hadrons.
We can now write the contributions of the spin $1/2$ baryon to the observed nucleon 
structure functions as convolutions \cite{KM}
%\begin{equation}
\begin{multline}
\frac{m_{N}^{2} q^{2} + (s_{N} \cdot q)^{2}}{(p_{N} \cdot q)^{2}} \delta g_{1}(x) - 
\gamma^{2} \delta g_{2}(x)  =    \\
\int_{x}^{y_{max}} \frac{dy}{y} \left[ B_{1}(y) g_{1}^{B} \left( \frac{x}{y}, Q^{2} \right) + 
B_{2}(y) g_{2}^{B} \left( \frac{x}{y}, Q^{2} \right) \right]
\label{eq:B12}
\end{multline}
%\end{equation}
where $y_{max}$ is the maximum allowable value of the momentum fraction 
$y = k \cdot q / p_{N} \cdot q$, which is usually 1, and the baryon momentum distributions are 
\bea
B_{1,2}(y) = \int_{0}^{(k_{\perp}^{2})_{max}} d \vec{k}_{\perp}^{2} \int_{0}^{2\pi} d \phi 
\frac{|\vec{p}_{N}|}{(2 \pi)^{3}} y \frac{\partial y^{\prime}}{\partial y} 
\sum_{\lambda_{B}, \lambda_{M}} |J_{NBM}|^{2} A_{1,2}.
\label{eq:Bcoeffs}
\eea
Here $y^{\prime}$ is the longitudinal momentum fraction
\bea
\vec{k} = \vec{k_{\perp}} + y^{\prime} \vec{p}_{N}
\eea
which, in the infinite momentum frame ($|\vec{p}_{N}| \ra \infty$), is related to $y$ by
\bea
y^{\prime} = \frac{y}{1 + \sqrt{1+ \gamma^{2}}} \left[ 1 + 
\sqrt{1 + \frac{\gamma^{2}}{y^{2}m_{N}^{2}}(\vec{k}_{\perp}^{2} + m_{B}^{2})} \right].
\eea
Most previous calculations of structure functions in the MCM have not taken into account the 
difference between the light-cone momentum fraction $y$ and the longitudinal momentum fraction 
$y^{\prime}$, as these are the same in the Bjorken limit. 
However at finite $Q^{2}$ the difference is not negligible.

The maximum transverse momentum squared is 
\bea
(k_{\perp}^{2})_{max} & = & \frac{m_{N}^{2}}{\gamma^{2}}(1 + \sqrt{1 + \gamma^{2}})
(1 - 2y + \sqrt{1 + \gamma^{2}}) - m_{B}^{2} \nonumber \\
& \ra & Q^{2} \frac{1-y}{x^{2}} - m_{B}^{2} \gg m_{N}^{2},
\eea
which at small $x$ is much larger than any momentum cut-off that is required for the vertex, 
so $(k_{\perp}^{2})_{max}$ may safely be taken to infinity.

From equation (\ref{eq:B12}) we see that the nucleon structure functions pick up contributions 
from both $g_{1}^{B}$ and $g_{2}^{B}$ of the baryon in the cloud. 
This occurs because the spin vector of the cloud baryon $s_{B}^{\mu}$ is not parallel to the 
initial nucleon spin vector $s_{N}^{\mu}$. 
So if the initial nucleon is longitudinally polarized, the baryon in the cloud will have 
both longitudinal and transverse spin components, and hence $g_{2}^{B}$ will give a finite 
contribution to $g_{1}^{N}$.
Similarly $g_{2}^{N}$ will get a contribution from $g_{1}^{B}$. 
As the `bare' $g_{2}^{N}$ structure functions are expected to be small, this kinematic contribution 
from the baryon - meson cloud could be a major portion of these structure functions.

Following reference \cite{KM} we write equation (\ref{eq:B12}) in terms of the nucleon and 
baryon helicities, $\lambda_{N}$ and $\lambda_{B}$ and the nucleon transverse spin vector 
$s_{N}^{\top}$ 
\begin{multline}
%\begin{split}
(\lambda_{N}^{2} - \frac{|s_{N}^{\top}|^{2}}{m_{N}^{2}} \gamma^{2}) \delta g_{1}(x) - 
\gamma^{2} \delta g_{2}(x)  =    \\ 
\sum_{\lambda_{B}} \lambda_{B} \int_{x}^{y_m} \frac{dy}{y} 
\left\{ [\lambda_{N}^{2} f_{1L}(y) + \frac{|s_{N}^{\top}|^{2}}{m_{N}^{2}} f_{1T}(y)] 
g_{1}^{B} \left( \frac{x}{y}, Q^{2} \right) + \right.  \\
 \left. [\lambda_{N}^{2} f_{2L}(y) + \frac{|s_{N}^{\top}|^{2}}{m_{N}^{2}} f_{2T}(y)] 
g_{2}^{B} \left( \frac{x}{y}, Q^{2} \right) \right\},
\label{eq:totconv}
%\end{split}
\end{multline}
where the momentum distributions $f_{1,2 \, L,T}(y)$ are similar (up to signs) to those given 
in equations (2.25 - 2.28) of reference \cite{KM} with $m_{V}$ and $\lambda_{V}$ replaced by 
$m_{B}$ and $\lambda_{B}$ respectively. We give these expressions in Appendix A below.

By combining the longitudinal $\lambda_{N} = 1\, (|s_{N}^{\top}|/m_{N} \equiv \tau_{N} =  0)$ 
with the transverse $\lambda_{N} = 0\, (\tau_{N}=1)$ amplitude in equation (\ref{eq:totconv}), 
and defining the functions
\bea
\Delta f_{1,2 \, L,T}^{1}(y) = f_{1,2 \, L,T}^{\lambda = +1}(y) - f_{1,2 \, L,T}^{\lambda = -1}(y)
\label{eq:deltaf}
\eea
we can obtain the separate contributions to the nucleon 
$g_{1}$ and $g_{2}$ structure functions:
\bea
\delta g_{1}(x, Q^{2}) & = & \frac{1}{1+\gamma^{2}} \int_{x}^{1} \frac{dy}{y} \sum_{i = 1,2} 
[\Delta f_{i L}^{1}(y) - \Delta f_{i T}^{1}(y)] g_{i}^{B} \left( \frac{x}{y}, Q^{2} \right) \\
\delta g_{2}(x, Q^{2}) & = & -\frac{1}{1+\gamma^{2}} \int_{x}^{1} \frac{dy}{y} \sum_{i = 1,2} 
\left[\Delta f_{i L}^{1}(y) + \frac{\Delta f_{i T}^{1}(y)}{\gamma^{2}} \right] 
g_{i}^{B} \left( \frac{x}{y}, Q^{2} \right).
\label{eq:dgs}
\eea

\subsection{Spin $1$ Mesons}

The analysis for spin 1 mesons exactly follows that above for spin 1/2 baryons, and was 
given by Kumano and Miyama \cite{KM}. 
The reason for this is that the most general form of the antisymmetric part of the meson 
tensor is the same for spin 1 mesons as for spin 1/2 baryons \cite{JM}.
Hence the results of the previous subsection can be directly translated to the vector 
meson case, simply by replacing $m_{B}$ and $\lambda_{B}$ by $m_{M}$ and $\lambda_{M}$ 
respectively, and replacing the baryon structure functions by meson structure functions.

Interestingly, the symmetric part of the meson tensor for spin 1 mesons contains two  
additional terms, which are both proportional to the structure function $b_{1}^{M}(x)$ 
at leading twist (via a generalized Callan-Gross relation). 
This would lead to a contribution to the nucleon structure function $F_{2}(x)$ coming 
from (for example) $b_{1}^{\rho}$, though it is expected that this will be rather small 
compared to the dominant MCM contribution coming from the pions in the cloud via 
$F_{2}^{\pi}$.

\subsection{Spin $3/2$ Baryons}

The number of independent Lorentz invariant structure functions for a spin $J$ hadron 
increases approximately linearly with $J$.
If $A^{J}_{h H, h^{\prime} H^{\prime}}$ are the imaginary part of the forward Compton helicity
amplitudes for 
$ \gamma_{h} + \mbox{hadron}^{J}_{H} \ra \gamma_{h^{\prime}} + \mbox{hadron}^{J}_{H^{\prime}}$, 
it can be seen that there are $6J + 2$ ($6J + 1$) independent amplitudes for $J$-integer 
(half-integer) satisfying P and T invariance and helicity conservation.
Of these, $2J$ ($2J+1$) amplitudes contribute to spin dependent scattering. 
Thus the general expression for the antisymmetric part of the hadronic tensor for a particle
of spin $J$ is \cite{JM}
\bea
W_{\mu \nu}^{A} & = & \sum_{L=1,3 \ldots}^{2J} i \frac{^{J}_{L}g_{1}}{\nu^{L}}
\epsilon^{\mu \nu \alpha \mu_{1}} \theta_{\mu_{1} \mu_{2} \ldots \mu_{L}} 
q_{\alpha} q^{\mu_{2}} \cdots q^{\mu_{L}} + \nonumber \\
& & \sum_{L=1,3 \ldots}^{2J} i \frac{^{J}_{L}g_{2}}{\nu^{L+1}} \epsilon^{\mu \nu \alpha \beta} 
p_{[\mu_{1}} \theta_{\beta ]  \mu_{2} \ldots \mu_{L}} 
q_{\alpha} q^{\mu_{1}} \cdots q^{\mu_{L}}.
\eea
In this expression $\theta^{\mu_{1} \mu_{2} \ldots \mu_{L}}$ is a completely symmetric, 
traceless pseudotensor.
It can be thought of as a generalized Pauli-Lubanski spin vector. 
For spin $1/2$ and spin $1$ only $\theta^{\mu}$ is non-vanishing, and it is proportional 
to the usual spin vector $s^{\mu}$. 
The structure functions $^{J}_{L}g_{1,2}$ are generalizations of the usual spin dependent 
structure functions.
At leading twist we have 
\bea
^{J}_{L}g_{1} & = & \sum_{H=-J}^{J} \langle J, H, J, -H | L, 0 \rangle q_{\ua}^{JH} \nonumber \\
^{J}_{L}g_{2} & = & 0,
\eea
and for $J=3/2$ in particular
\bea
^{\frac{3}{2}}_{1}g_{1} & = & \frac{1}{\sqrt{20}}
(3q_{\ua}^{\frac{3}{2} \frac{3}{2}} - 3q_{\da}^{\frac{3}{2} \frac{3}{2}} + 
q_{\ua}^{\frac{3}{2} \frac{1}{2}} - q_{\da}^{\frac{3}{2} \frac{1}{2}}) \nonumber \\
^{\frac{3}{2}}_{3}g_{1} & = & \frac{1}{\sqrt{20}}
(q_{\ua}^{\frac{3}{2} \frac{3}{2}} - q_{\da}^{\frac{3}{2} \frac{3}{2}} - 
3q_{\ua}^{\frac{3}{2} \frac{1}{2}} + 3q_{\da}^{\frac{3}{2} \frac{1}{2}}).
\eea
The polarization vectors for a spin $3/2$ particle have a spinor nature \cite{FG} which 
slightly complicates the expression for the Pauli-Lubanski vector
\bea
s^{\mu}(\frac{3}{2}, \lambda) = \frac{3}{2} i \epsilon^{\mu \rho \sigma \tau} 
\mbox{Tr}[ \bar{E}_{\rho}({\lambda}) E_{\sigma}({\lambda})]p_{\tau}
\eea
where $\lambda$ can take the values $\pm 1/2, \; \pm 3/2$, and the trace is taken over 
spinor indices. 
We have the normalization $s^{2} = -\lambda^{2}M^{2}$.

As in the spin $1/2$ and spin $1$ cases we can take $\theta^{\mu} \propto  s^{\mu}$. 
We also require $\theta^{\mu \nu \rho}$ in the spin $3/2$ case. We take the traceless 
combination of symmetric pseudotensors
\bea
\theta^{\mu \nu \rho} \propto \frac{1}{\lambda^{2}} s^{\mu} s^{\nu} s^{\rho} + 
{\cal S} (s^{\mu} p^{\nu} p^{\rho})
\eea
where $\cal S$ symmetrises over the Lorentz indices. After some work we obtain
\bea
W_{\mu \nu}^{A} & = & i \epsilon_{\mu \nu \rho \sigma} q^{\rho} \left[ \frac{s^{\sigma}}{\nu} 
\left(\frac{2}{\sqrt{5}} {}^{\frac{3}{2}}_{1}g_{1} - \frac{2 \omega^{2}}{2\sqrt{5}} {}^{\frac{3}{2}}_{3}g_{1} \right) 
 - \frac{4}{3\sqrt{5}} \frac{p^{\sigma} s\cdot q}{\nu^{2}} {}^{\frac{3}{2}}_{3}g_{1} + \right. \nonumber \\
& & \phantom{i \epsilon_{\mu \nu \rho \sigma} q^{\rho} \left[ \right.}\left. 
\left(\frac{s^{\sigma}}{\nu} -  \frac{p^{\sigma} s\cdot q}{\nu^{2}} \right) 
\left( \frac{2}{\sqrt{5}} {}^{\frac{3}{2}}_{1}g_{2} - \frac{2 \omega^{2}}{2\sqrt{5}} {}^{\frac{3}{2}}_{3}g_{2} \right) \right]
\eea
where 
\bea
\omega^{2} = 1 + \frac{(s \cdot q)^{2}}{\lambda^{2} \nu^{2}}
\eea
which goes to $2$ in the Bjorken limit.

We can now follow the same steps as for spin $1/2$ and spin $1$ hadrons to obtain the MCM 
contributions to the spin dependent structure functions of the nucleon.
In this case the generalizations of the coefficients $A_{i}$ and $B_{i}$ above depend on 
the helicity of the struck baryon, so it is useful to rewrite the spin $3/2$ structure 
functions as $g_{i}^{JH}$ which depend only on one helicity state.
We have
\bea
g_{i}^{\frac{3}{2} \frac{3}{2}} & = & \frac{3}{2\sqrt{5}} {}^{\frac{3}{2}}_{1}g_{i} + 
\frac{1}{2\sqrt{5}} {}^{\frac{3}{2}}_{3}g_{i} \nonumber \\
g_{i}^{\frac{3}{2} \frac{1}{2}} & = & \frac{1}{2\sqrt{5}} {}^{\frac{3}{2}}_{1}g_{i} - 
\frac{3}{2\sqrt{5}} {}^{\frac{3}{2}}_{3}g_{i}.
\eea
This gives 
\begin{multline}
\tilde{P}^{\mu \nu} W_{\mu \nu}^{(3/2) A}(k, s_{B}, \lambda, q) =   \\
\frac{m_{N}}{m_{B}} [A_{1}^{\frac{3}{2}}g_{1}^{\frac{3}{2} \frac{3}{2}}(k,q) + 
A_{1}^{\frac{1}{2}}g_{1}^{\frac{3}{2} \frac{1}{2}}(k,q) + 
A_{2}^{\frac{3}{2}}g_{2}^{\frac{3}{2} \frac{3}{2}}(k,q) + 
A_{2}^{\frac{1}{2}}g_{2}^{\frac{3}{2} \frac{1}{2}}(k,q)]
\end{multline}
where the coefficients are linear combinations of $A_{1}$ and $A_{2}$ from equation 
(\ref{eq:Acoeffs2})
\bea
A_{1}^{\frac{3}{2}} &  = & \frac{14 - 2 \omega^{2}}{15} A_{1} + \frac{4}{15} A_{2} \nonumber \\
A_{1}^{\frac{1}{2}} &  = & \frac{6 + 2 \omega^{2}}{5} A_{1} - \frac{4}{5} A_{2} \nonumber \\
A_{2}^{\frac{3}{2}} &  = & \frac{18 - 2 \omega^{2}}{15} A_{2}  \nonumber \\
A_{2}^{\frac{1}{2}} &  = & \frac{2 + 2 \omega^{2}}{15} A_{2}
\eea
and 
\bea
\omega^{2} = 1 + \left[1 - \frac{m_{H}^{2}}{yy^{\prime}m_{N}^{2}}(1 - \sqrt{1 + \gamma^{2}}) \right]^{2}.
\eea
Now doing the required integrations (details can be found in Appendix A) we end up with a 
result similar to equation (\ref{eq:dgs}):
\bea
\delta g_{1}(x, Q^{2}) & = & \frac{1}{1+\gamma^{2}} \int_{x}^{1} \frac{dy}{y} 
\sum_{i = 1,2} \sum_{h = \frac{1}{2},\frac{3}{2}} 
[\Delta f_{i L}^{\frac{3}{2} h}(y) - \Delta f_{i T}^{\frac{3}{2} h}(y)] 
g_{i}^{\frac{3}{2} h} \left( \frac{x}{y}, Q^{2} \right) \\
\delta g_{2}(x, Q^{2}) & = & -\frac{1}{1+\gamma^{2}} \int_{x}^{1} \frac{dy}{y} 
\sum_{i = 1,2} \sum_{h = \frac{1}{2},\frac{3}{2}}
\left[\Delta f_{i L}^{\frac{3}{2} h}(y) + \frac{\Delta f_{i T}^{\frac{3}{2} h}(y)}{\gamma^{2}} \right] 
g_{i}^{\frac{3}{2} h} \left( \frac{x}{y}, Q^{2} \right).
\eea

\section{Interference contributions}

\begin{figure}[htb]
\begin{center}
\includegraphics[width= 8 cm]{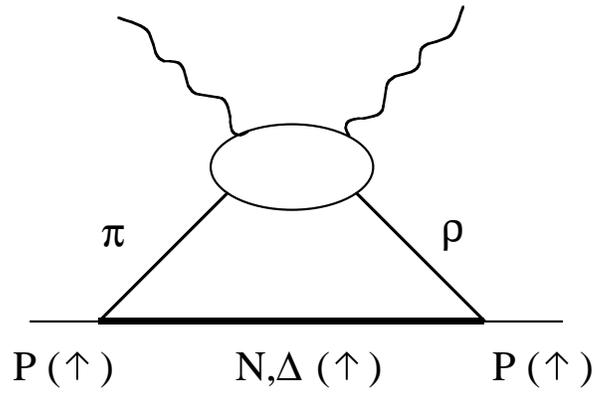}
%\vskip -0.5cm
\caption{Interference between $\pi$ and $\rho$ mesons.}
\label{fig:Int}
\end{center}
\end{figure}

In polarised DIS, we can consider the possibility of interference terms between intermediate 
states containing different hadrons.
This possibility is allowed in polarised DIS as the observed spin dependent structure 
functions do not contribute to the total $\gamma^{*}N$ cross section, but only to 
$\Delta \sigma = \sigma_{\frac{3}{2}} - \sigma_{\frac{1}{2}}$ or, equivalently, to 
$\sigma_{I}$ the cross section associated with interference between transverse and 
longitudinal polarisations of the virtual photon \cite{Roberts}.
Previous authors have considered interference between $\pi$ and $\rho$ mesons 
\cite{BK, FCaoS_plzdsea}, $\pi$ and $\sigma$ mesons \cite{FSW}, $K$ and $K^{*}$ mesons 
\cite{FCaoS_KKstar}, and $N$ and $\Delta$ baryons \cite{ScT, CBorosT, FCaoS_ps2}. 
We show an example of an interference term in figure~\ref{fig:Int}.

The interference terms between mesons of different helicity are particularly interesting 
as they appear to offer a mechanism whereby angular momentum in the cloud can be directly 
coupled to quarks, possibly giving rise to large sea quark polarizations \cite{BK}.
However care needs to be taken over which interference terms can actually contribute to 
the observed structure functions. 
Let $\tilde{A}_{h H, h^{\prime} H^{*}}$ be the imaginary part of the forward helicity 
amplitude for the interference term
$ \gamma_{h} + \mbox{hadron1}_{H} \ra \gamma_{h^{\prime}} + \mbox{hadron2}_{H^{*}}$. 
In the Bjorken limit quark helicity is conserved, which implies that the only amplitudes that 
contribute to $\Delta \sigma$ are in the combinations
$(\tilde{A}_{1 H, 1 H^{*}=H} - \tilde{A}_{-1 H, -1 H^{*}=H})$.
Also the generalization of the Callen-Gross relation gives 
$\tilde{A}_{0 H, 0 H^{*}=H} \ra 0$.
These two results imply that for (pseudo)scalar mesons interfering  with (pseudo)vector mesons 
only combinations of amplitudes  
$(\tilde{A}_{1 0, 1 0^{*}} - \tilde{A}_{-1 0, -1 0^{*}})$
can contribute to the structure function.
However this combination will be zero by parity invariance, meaning that interference between 
mesons cannot contribute to the leading twist structure functions.
Amplitudes like $\tilde{A}_{1 0, 0 H^{*}=1}$ can contribute to the interference cross section
$\sigma_{I}$ between transverse and longitudinal photon polarisations at higher twist.

In the case of interference between $N$ and $\Delta$ baryons this contribution involves the 
combination of amplitudes
$(\tilde{A}_{1 \frac{1}{2}, 1 H^{*}=\frac{1}{2}} - \tilde{A}_{-1 \frac{1}{2}, -1 H^{*}=\frac{1}{2}})$ 
which need not vanish. 

We can write the contribution of interference terms to the nucleon tensor $W_{\mu \nu}$, where 
the two particles that interfere are labelled $X$ and $Y$ and the spectator hadron is labelled 
by $S$  
\begin{multline}
\delta W_{\mu \nu}^{XY}(p_{N}, s_{N}, q) = \\
\int \frac{d^{3}p_{S}}{(2\pi)^{3}} \frac{2\sqrt{m_{X}m_{Y}}m_{S}}{E_{S}}
\sum_{\lambda_{X}, \lambda_{Y}, \lambda_{S}} 
\left[ J_{NXS} J^{*}_{NYS} W_{\mu \nu}^{X \ra Y}(k_{X}, s_{X}, k_{Y}, s_{Y}, q) \right. \\
+ \left. J_{NXS} J^{*}_{NYS} W_{\mu \nu}^{Y \ra X}(k_{Y}, s_{Y}, k_{X}, s_{X}, q) \right]
\label{eq:Winf}
\end{multline}
where the interference tensor is given by
\begin{multline}
W_{\mu \nu}^{X \ra Y}(k_{X}, s_{X}, k_{Y}, s_{Y}, q) = \\ 
\frac{1}{4\pi \sqrt{m_{X}m_{Y}}} \sum_{X', Y'} \delta^{4}(p_{X'}^{2} - m_{X'}^{2}) 
\delta^{4}(p_{Y'}^{2} - m_{Y'}^{2}) \langle k_{X}, s_{X} | J_{\mu} | X' \rangle 
\langle Y' | J_{\nu}^{\dagger} | k_{Y}, s_{Y} \rangle.
\end{multline}
We see that the interference tensors in equation (\ref{eq:Winf})are related by
\bea
W_{\mu \nu}^{Y \ra X}(k_{Y}, s_{Y}, k_{X}, s_{X}, q) = 
\left[W_{\nu \mu}^{X \ra Y}(k_{X}, s_{X}, k_{Y}, s_{Y}, q) \right]^{*}.
\eea
This enables us to write the contribution to the nucleon tensor as 
\begin{multline}
\delta W_{\mu \nu}^{XY}(p_{N}, s_{N}, q) = \\
\int \frac{d^{3}p_{S}}{(2\pi)^{3}} \frac{2\sqrt{m_{X}m_{Y}}m_{S}}{E_{S}}
\sum_{\lambda_{X}, \lambda_{Y}, \lambda_{S}} 2 { \cal R} \left[ J_{NXS} J^{*}_{NYS} \right]
W_{\mu \nu}^{X \ra Y}(k_{X}, s_{X}, k_{Y}, s_{Y}, q).
\label{eq:Winf2}
\end{multline}

In time ordered perturbation theory (TOPT), the 3-vectors of the interfering particles 
$\vec{k}_{X, Y}$  will be identical, however their energies are not, as both particles are 
on-shell.
We introduce two momentum fractions
\bea
y_{X, Y} = \frac{k_{X,Y}\cdot q}{p_{N}\cdot q}, 
\eea
noting that the longitudinal momentum fraction $y^{\prime}$ is the same for both hadrons.
If we define 
\bea
k^{\mu} = \frac{1}{2}(k_{X}^{\mu} + k_{Y}^{\mu}), & \;\; &
\delta k^{\mu} = \frac{1}{2}(k_{X}^{\mu} - k_{Y}^{\mu}) \\
y = \frac{1}{2}(y_{X} + y_{Y}), & \;\; &
\delta y = \frac{1}{2}(y_{X} - y_{Y}) \\
s^{\mu} = \frac{1}{2}(s_{X}^{\mu} + s_{Y}^{\mu}), & \;\; &
\delta s^{\mu} = \frac{1}{2}(s_{X}^{\mu} - s_{Y}^{\mu}) 
\eea
then in the Bjorken limit we have $\delta k \ra 0$, $\delta y \ra 0$ and $\delta s \ra 0$.

We can now write the most general form of the antisymmetric tensor for the interference term 
\bea
\delta W_{\mu \nu}^{(A) \, XY} & = & i\epsilon_{\mu \nu \rho \sigma} q^{\rho} 
\left[ \frac{s^{\sigma}}{k \cdot q} g_{1}^{XY} + 
\left(\frac{s^{\sigma}}{k \cdot q} - \frac{s \cdot q}{(k \cdot q)^{2}} k^{\sigma}\right) g_{2}^{XY} \right. \nonumber \\
&& \phantom{i\epsilon_{\mu \nu \rho \sigma} q^{\rho} \left[ \right.}
\left. +  \frac{\delta s^{\sigma}}{k \cdot q} \tilde{g}_{1}^{XY} + 
\frac{s \cdot q}{(k \cdot q)^{2}} \delta k^{\sigma} \tilde{g}_{2}^{XY} \right] \nonumber \\
&& + i \frac{\tilde{k}_{X}^{\mu} \tilde{k}_{Y}^{\nu} - \tilde{k}_{X}^{\nu} \tilde{k}_{Y}^{\mu}}{2k \cdot q}
\tilde{a}_{1}^{XY} + 
i \frac{\tilde{s}_{X}^{\mu} \tilde{s}_{Y}^{\nu} - \tilde{s}_{X}^{\nu} \tilde{s}_{Y}^{\mu}}{2k \cdot q}
\tilde{a}_{2}^{XY}, 
\eea
where we denote $\tilde{v}^{\mu} = (v^{\mu} - u \cdot q q^{\mu}/q^{2})$ for any four vector $v$.
This includes four possible new interference `structure functions'. 
All of these new terms arise from our use of TOPT, and would vanish in a covariant formulation. 
However the price to be paid in the covariant formulation is that we would have to use 
structure functions of off-shell hadrons, which are difficult to define and measure.
As $\delta k \approx \left( \frac{m_{Y}^{2} - m_{X}^{2}}{4y^{\prime}p}, \vec{0}\right)$ 
and 
$\delta s \approx \lambda \left(0, \frac{m_{Y}^{2} - m_{X}^{2}}{4(y^{\prime}p)^{2}}\vec{k}_{\perp}, 
\frac{m_{Y}^{2} - m_{X}^{2}}{4y^{\prime}p} \right)$, 
the contributions from $\tilde{g}_{1,2}^{XY}$ will be kinematically suppressed, 
and look like higher twist corrections to the observed $g_{1}$ and $g_{2}$.

The two structure functions $\tilde{a}_{1}^{XY}$ and $\tilde{a}_{2}^{XY}$ involve antisymmetric 
combinations of hadron four vectors and are independent of polarisation. 
These terms do not give any contribution to the nucleon structure functions because when the 
projector $\tilde{P}^{\mu \nu}$ is applied to them we obtain a coefficient $A_{i}^{XY}$ which 
is proportional to $\sin \phi$.
Integration over $\phi$ then results in these terms being zero.
This agrees with our earlier observation that interference involving (pseudo)scalar mesons 
and (pseudo)vector or scalar mesons does not contribute at leading twist, as this violates 
parity invariance.

In a similar fashion to our procedure in the previous section, we multiply the antisymmetric 
interference tensor by the projector 
$\tilde{P}^{\mu \nu} = (m_{N}/\sqrt{m_{X}m_{Y}}) P^{\mu \nu}$, which gives  
\bea
\tilde{P}^{\mu \nu} \delta W_{\mu \nu}^{(A) \, XY} & = & 
\frac{m_{N}}{\sqrt{m_{X}m_{Y}}} 
\left[ A_{1}^{XY} g_{1}^{XY} + \tilde{A}_{1}^{XY} \tilde{g}_{1}^{XY} +
A_{2}^{XY} g_{2}^{XY} + \tilde{A}_{2}^{XY} \tilde{g}_{2}^{XY} \right],
\label{eq:Acoeffsinf}
\eea
where 
\bea
 A_{1}^{XY} & = & \frac{1}{p_{N}\cdot q \, k\cdot q} 
(s_{N}\cdot q \, s \cdot q - q^{2} s_{N}\cdot s)   \\
\tilde{A}_{1}^{XY} & = & \frac{1}{p_{N}\cdot q \, k\cdot q} 
(s_{N}\cdot q \, \delta s \cdot q - q^{2} s_{N}\cdot \delta s) \\
A_{2}^{XY} & = & \frac{q^{2}}{p_{N}\cdot q \, (k\cdot q)^{2}} 
(s_{N}\cdot k \, s \cdot q - k \cdot q \, s_{N}\cdot s)  \\
\tilde{A}_{2}^{XY} & = & \frac{s \cdot q}{p_{N}\cdot q \, (k\cdot q)^{2}} 
(s_{N}\cdot q \, \delta k \cdot q - q^{2} s_{N}\cdot \delta k). 
\eea
We observe that $A_{1}^{XY}$ and $A_{2}^{XY}$ are the same 
(up to mass factors) as the coefficients given in equations (\ref{eq:Acoeffs1}) and 
(\ref{eq:Acoeffs2}) for spin 1/2 baryons in the cloud. 
For the other two coefficients we find 
\bea
\tilde{A}_{1}^{XY} & = & \lambda \lambda_{N} 
\frac{m_{X}^{2}-m_{Y}^{2}}{4y y^{\prime}m_{N}^{2}}(1 - \sqrt{1+\gamma^{2}}) \\
\tilde{A}_{2}^{XY} & = & -\tilde{A}_{1}^{XY} 
\left(1 - \frac{m_{X}^{2}-m_{Y}^{2}}{y y^{\prime}m_{N}^{2}}(1 - \sqrt{1+\gamma^{2}}) \right)
\eea
which both vanish in the Bjorken limit.

In the expression for the interference tensor (equation (\ref{eq:Winf2}))we can write the part of 
the integrand that depends on the vertices as a sum of polarization independent plus longitudinal 
and transverse terms
\bea
\frac{2\sqrt{m_{X}m_{Y}}m_{S}}{(2\pi)^{3} E_{S}} \sum_{\lambda_{S}} 
2 { \cal R} \left[ J_{NXS} J^{*}_{NYS} \right]
= C_{0}^{\lambda}(\phi) + \lambda_{N} C_{L}^{\lambda}(\phi) + \tau_{N}  C_{T}^{\lambda}(\phi)
\label{eq:ccoeffs}
\eea 
where $\phi$ is the angle between $k_{\perp}$ and $\vec{s}_{N}^{\top}$, 
$\tau_{N} = |\vec{s}_{N}^{\top}|/m_{N}$ and $\lambda$ labels the helicity of the struck hadron.
If we consider the case of interference between a pion and a rho with helicity $\pm1$, we find
that $C_0$ is zero while $C_L$ and $C_T$ are not. When combined with the appropriate $A^{XY}_i$ 
coefficients and integrated over $\phi$ we find that these contributions are zero.
Hence there are no contributions to the interference tensor from interference between $\pi$ and
$\rho$ mesons.
Similar conclusions can also be drawn about any contributions from interference between 
$K$ and $K^{*}$ mesons.
Interference between $N$ and $\Delta$ baryons also appears to be suppressed.
We find that when the spectator meson is a pion the coefficients $C_{0}, C_{L}$ and $C_{T}$ 
are non-zero, however their angular dependance is proportional to terms like $\cos \phi$,  $\sin \phi$
or $\cos 2\phi$, all of which again integrate to zero when combined with the appropriate
coefficients.
Details of this calculation are in Appendix B below.
In the case of the spectator meson being a $\rho$ meson the coefficients are very difficult to 
calculate because of the complicated gamma structure of the two vertices. 
However this contribution is already greatly suppressed because of the small probability of the 
$|\Delta \rho \rangle$ state.
Thus interference contributions to the polarized structure functions are mostly zero or very 
small in the meson cloud model, and we shall henceforth neglect them.

Our conclusions regarding the contributions of interference terms are different from those of earlier 
authors who considered these terms
 \cite{ScT, BK, CBorosT, FCaoS_plzdsea, FSW, FCaoS_KKstar, FCaoS_ps2}.
These earlier works generally calculated interference terms by considering separately 
$J_{NXS}$ and $J_{NYS}$ as given in appendix B of reference \cite{HHoltmannSS}, or 
similar, which are worked out from considering the direct processes. 
However it appears that these calculations of the vertex factors have not followed the same, 
or consistent, phase conventions when considering different vertices.
This is not important when considering direct processes, as all terms are proportional to 
$|J|^{2}$, but for interference terms involving $J_{NXS}J^{*}_{NYS}$, any change in relative 
phase between the two vertices renders the calculation meaningless.
In this work we have not calculated the two vertices separately, but considered the complete 
interference process.
The advantage of this is that the two amplitudes for  the processes $XS \ra YS$ and $YS \ra XS$ 
must be added. 
As these two amplitudes are  conjugate, the result must be real, which gives a check 
that the phase factors have been correctly accounted for.
More details are given in Appendix B.

\section{Spin dependent momentum distributions of mesons and baryons in the cloud}

We now turn to the calculation of the various momentum distributions 
$\Delta f(y)$ of the components of the cloud.
In general these distributions are of the form (up to kinematic factors given in the 
previous section and Appendix A)
\bea
\Delta f_{BM} \sim \int_{0}^{(k_{\perp}^{2})_{m}} dk_{\perp}^{2}
J_{NBM}^{\lambda_{1}\lambda_{2}}(y, k_{\perp}^{2}) 
(J_{NBM}^{\lambda_{1}\lambda_{2}}(y, k_{\perp}^{2}))^{\dagger}.
\eea
%where we have allowed for the possibility of interference terms.
In this case $J_{NBM}$ is the nucleon-baryon-meson vertex function multiplied by the 
propagator of the struck component of the cloud {\it i.e.}
\bea
J_{NBM}^{\lambda_{1}\lambda_{2}} \propto 
\frac{V(\vec{p}, \uparrow; \vec{k}, \lambda_{1}, \vec{p^{\prime}}, \lambda_{2})}
{E_{N} - E_{M} - E_{B}} \label{eq:JNMB}
\eea
which is the amplitude that a nucleon with momentum $\vec{p}$ and helicity $+1/2$ is
found in a meson cloud Fock state where the struck hadron has momentum $\vec{k}$ and 
helicity $\lambda_{1}$ and the spectator hadrom has momentum $\vec{p^{\prime}}$ and 
helicity $\lambda_{2}$. Note that we have explicitly used time-ordered perturbation 
theory (TOPT) to write the propagator of the struck particle being proportional to 
the energy denominator in this expression.

In the infinite momentum frame (IMF), $p = |\vec{p}\,| \rightarrow \infty$, Drell, Levy and Yan 
\cite{DLY} (building on earlier work by Weinberg \cite{Wein}) showed that contributions from 
Fock states containing anti-particles vanish and also that only the contributions with forward 
moving ($y \geq 0)$ particles contribute. 
As we saw earlier, all relevant momenta can be expressed in terms of $y$, which we take as 
the longitudinal momentum proportion carried by the meson, and $\vec{k_{\perp}}$.
The amplitude is now proportional to 
\bea
\frac{V^{\lambda_{1} \lambda_{2}}_{IMF}(y, k_{\perp}^{2})}
{m_{N}^{2} - M_{BM}^{2}(y, k_{\perp}^{2})}
\eea
where 
\bea
M_{BM}^{2}(y, k_{\perp}^{2}) = \frac{m_{M}^{2} + k_{\perp}^{2}}{y} + 
\frac{m_{B}^{2} + k_{\perp}^{2}}{1-y}
\eea
is the invariant mass squared of the Fock state.

Using TOPT guarantees that, for a given $|BM\rangle$ component of the cloud, the probability 
of finding the meson $M$ with longitudinal momentum fraction $y$ is equal to that of finding the 
baryon $B$ with longitudinal momentum fraction $1-y$. 
This is not necessarily true in a covariant approach \cite{HHoltmannSS}, which leads to 
non-conservation of charge and momentum. 
However a problem arises in TOPT when the vertex contains derivative coupling between 
fields (e.g. the usual pseudovector $\rho NN$ vertex contains a term proportional to 
$\bar{\psi} \sigma_{\mu \nu} (\partial^{\mu}\theta^{\nu}-\partial^{\nu}\theta^{\mu})\psi$ 
where $\psi$ is the nucleon field and $\theta$ is the vector field), as these terms introduce 
off-shell dependence into the vertex function which is not suppressed in the IMF.
This leads to two possible choices for the meson energy: A) the on-shell meson energy 
$E_{M} = \sqrt{m_{M}^{2} + k^{2}}$, or B) the off-shell meson energy, i.e. the  
difference between baryon energies $E_{N} - E_{B}$. 
While the second choice may be more `natural' in that the vertex structure is only due 
to baryonic currents \cite{WT94, SpethT}, the first appears more compatible with TOPT in that 
the meson remains on-shell. 
In practice a number of authors \cite{FS, KM} have used both prescriptions, and treated them 
as two different models. 

We can gain some insight into the choice of meson energy if we recall that TOPT in the 
IMF is equivalent to light-cone perturbation theory (LCPT) \cite{DLY, KS, LB}. 
In LCPT it is important to be aware of the light-cone singularities in the particle 
propagators. 
For spin zero particles the Klein-Gordon propagator (in light-cone co-ordinates) is \cite{KS}
\bea
\Delta_{F}(x) = \frac{-i}{(2\pi)^{3}} \int d^{2}p_{\perp} \int_{0}^{\infty} \frac{dp^{+}}{2p^{+}}
\left[ \Theta(x^{+}) e^{-ip \cdot x} + \Theta(-x^{+}) e^{ip \cdot x} \right].
\eea
The singularity at $p^{+} = 0$ does not affect the light-cone behaviour of the propagator, 
which is governed by the light-cone discontinuities $\Theta(\pm x^{+})$. 
For particles of higher spin, the propagators all pick up terms proportional to 
\bea
\delta(x^{+}) \frac{1}{(2\pi)^{3}}\int d^{2}p_{\perp} \int_{-\infty}^{\infty} \frac{dp^{+}}{2p^{+}} 
\exp[-i(p^{-}x^{-} - \vec{p_{\perp}} \cdot \vec{x_{\perp}})]
\eea
in addition to the terms proportional to $\Theta(\pm x^{+})$.
The term proportional to $\delta(x^{+})$ is an instantaneous part of the propagator.
This term can be absorbed into the regular propagator by replacing in the numerator of the 
diagrams in which the particle propagates over a single time interval, the momentum $p$ 
associated with the line by 
\bea
\tilde{p} = \left(p^{+}, \sum_{inc}p^{-} - \sideset{}{^\prime}\sum_{int}p^{-} ,\vec{p_{\perp}} \right)
\eea
where $\sum_{inc}$ sums over all the initial particles in the diagram and $\sum_{int}^{\prime}$ 
sums over all the particles in the intermediate state {\em except} the particle of interest 
\cite{LB}. 
Returning to TOPT in the IMF, we see that this is equivalent to choice B for the meson 
energy, for non-scalar mesons.
For spin-zero mesons there are no terms corresponding to instantaneous propagation (in light-cone
co-ordinates), and the propagator is not adjusted.
Hence for scalar and pseudoscalar mesons the correct meson energy is choice A, i.e. the mesons 
are always treated as on-shell.

The vertex function $V_{IMF}^{\lambda\lambda^\prime}(y,k_\perp^2)$ is calculated from 
the effective interaction Lagrangians (see Appendix B) which are usually employed in
the meson exchange models \cite{RMachleidtHE}.
Phenomenological vertex form factors $G_{BM}(y,k_\perp^2)$ are also introduced into equation 
(\ref{eq:JNMB}) to describe the unknown dynamics of the fluctuation $N\ra BM$ arising from 
the extended structure of hadrons. 
We use the exponential form factor
\bea
G_{BM}(y,k_\perp^2)={\rm exp} \left[\frac{m_N^2-m_{BM}(y,k_\perp^2)}
{2\Lambda_{BM}^{2}}\right],
\label{eq:GBM}
\eea
with $\Lambda_{BM}$ being a cut-off parameter, which is well defined in the model and provides 
a cut-off in both $t$ and $u$ (the four momentum squared of the intermediate baryon).
The form factor satisfies the relation
$G_{BM}(y,k_\perp^2)=G_{MB}(1-y,k_\perp^2)$. 
Using form factors introduces new parameters $\{ \Lambda_{BM} \}$ into any calculation 
using the MCM, with each Fock state having (in principle) its own cut-off. 
However the J\"{u}lich group \cite{HHoltmannSS} and Zoller \cite{Zoller} used high-energy 
particle production data to determine all the $\Lambda_{BM}$ of interest, and found that 
the data could be described by two parameters: $\Lambda_{1}$ for octet baryons and pseudoscalar 
and vector mesons, and $\Lambda_{2}$ for decuplet baryons. 
The upper limits for these two parameters were determined to be about 1 GeV, which is fairly 
soft, and gives the probability of all Fock states totalling to about 40\%. 
Melnitchouk, Speth and Thomas \cite{WMelnitchoukST} found a good fit to both the violation of 
the Gottfried sum rule \cite{NMC} and the observed ratio of $\bar{d(x)} / \bar{u}(x)$ 
from the E866 experiment \cite{E866} using values $\Lambda_{1} = 0.80$ GeV and 
$\Lambda_{2} = 1.00$ GeV, and we shall use these values of the cut-offs in this work. 

\begin{figure}[htb]
\begin{center}
\includegraphics[width=15 cm]{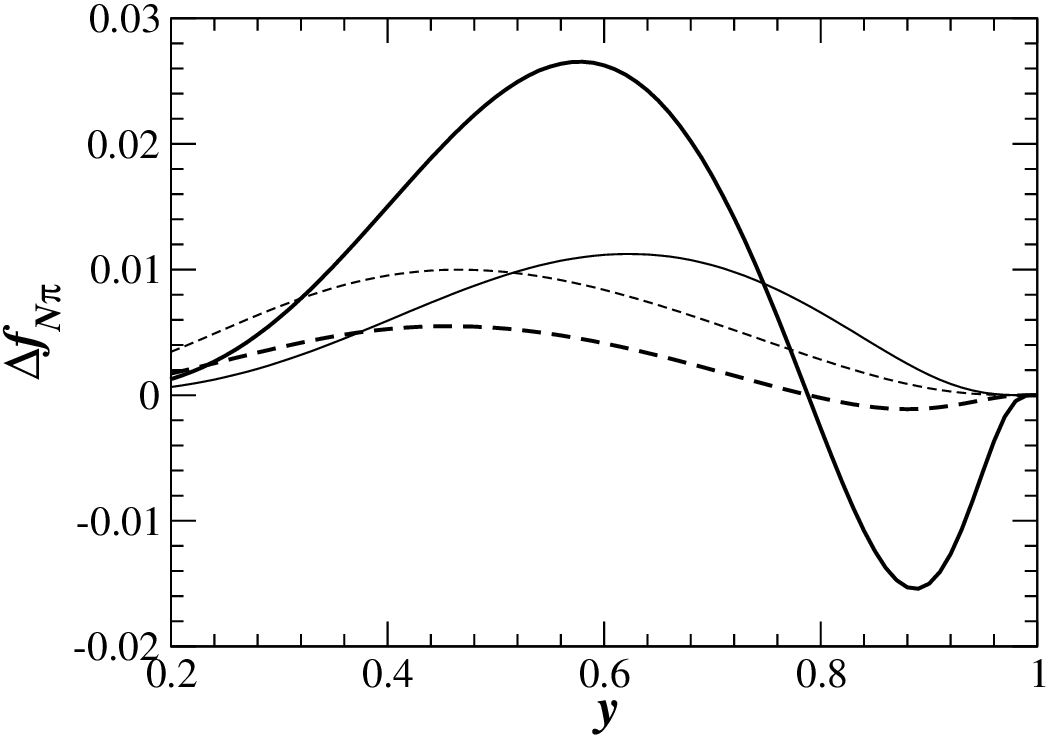}
\vskip -3.0cm
\caption{Fluctuation functions for $N \ra N \pi$ with $N$ being struck.
  The thick solid and dashed curves
  are for $\Delta f_{1L}$ and  $\Delta f_{2L}$, respectively. 
  The thin solid and dashed curves stand for
  $10 \, \Delta f_{1T}$ and $100 \, \Delta f_{2T}$, respectively.} 
\label{fig:DfNpi}
\end{center}
\end{figure}

\begin{figure}[htb]
\begin{center}
\includegraphics[width=15 cm]{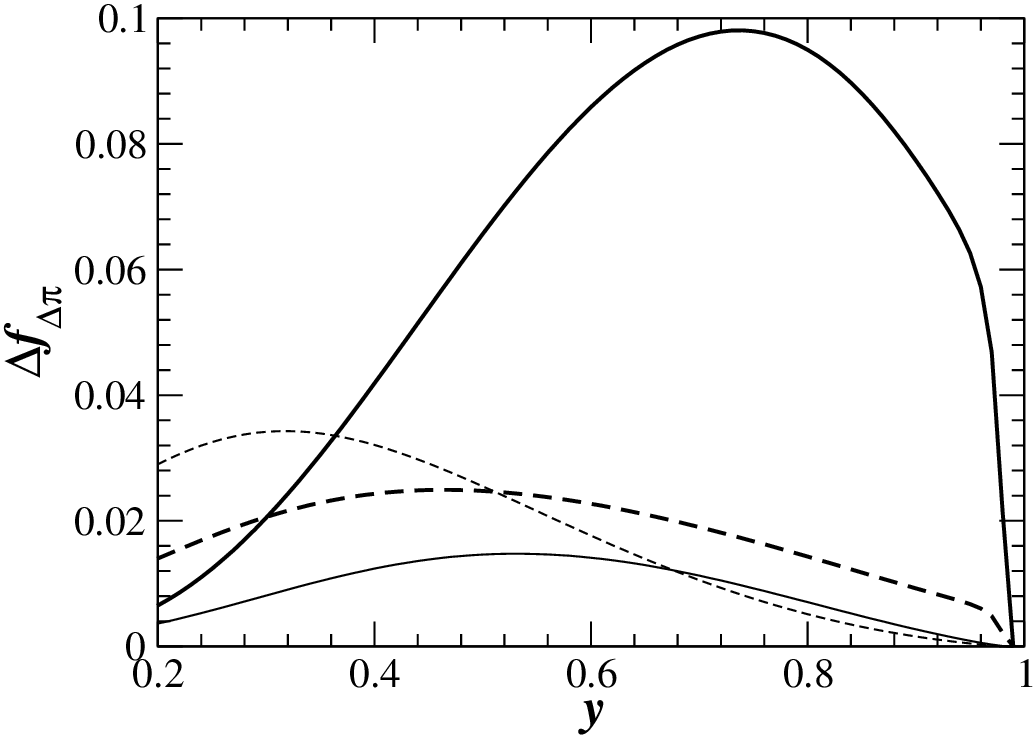}
\vskip -3.0cm
\caption{Fluctuation functions for $N \ra \Delta \pi$ with $\Delta$ being struck.
  The thick solid and dashed curves
  are for $\Delta f_{1L}$ and  $\Delta f_{2L}$, respectively. The thin solid and dashed curves stand for
  $10 \, \Delta f_{1T}$ and $100 \, \Delta f_{2T}$, respectively.
  $\Delta f_{1(2)L(T)}=\Delta f_{1(2)L(T)}^{\frac{3}{2}\frac{1}{2}}+
  3 \Delta f_{1(2)L(T)}^{\frac{3}{2}\frac{3}{2}}$.}
\label{fig:DfDpi}
\end{center}
\end{figure}

\begin{figure}[htb]
\begin{center}
\includegraphics[width=15 cm]{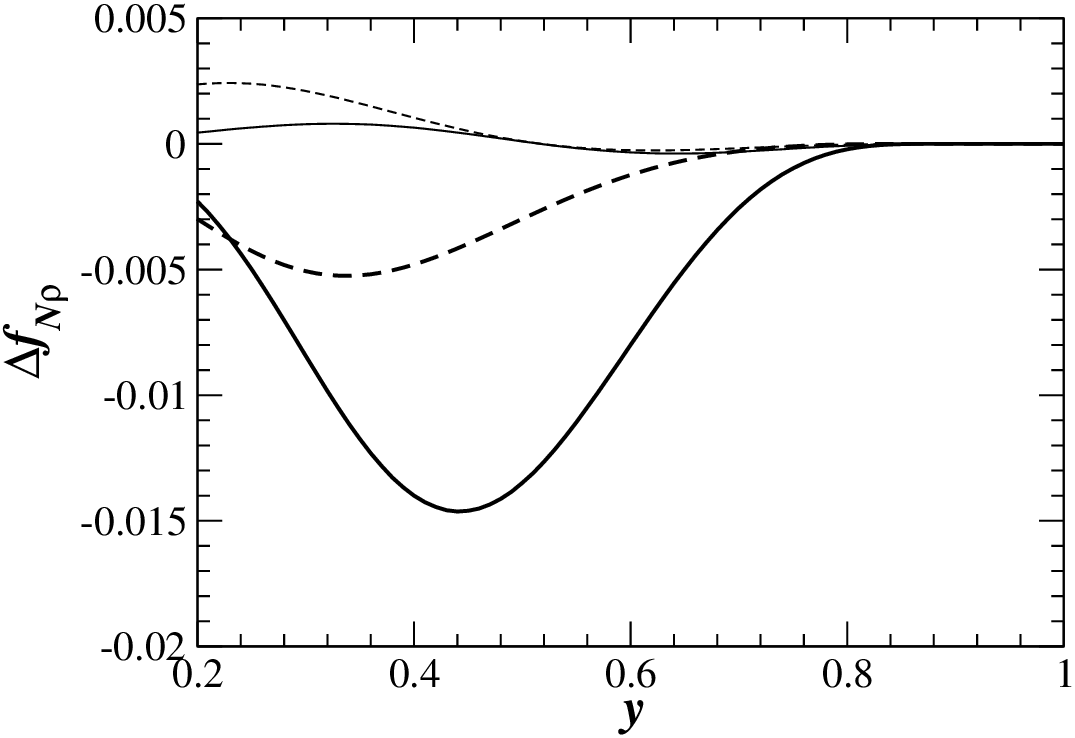}
\vskip -3.0cm
\caption{Fluctuation functions for $N \ra N \rho$ with $N$ being struck.
  The thick solid and dashed curves
  are for $\Delta f_{1L}$ and  $\Delta f_{2L}$, respectively. 
  The thin solid and dashed curves stand for
  $10 \, \Delta f_{1T}$ and $100 \, \Delta f_{2T}$, respectively.} 
\label{fig:DfNrho}
\end{center}
\end{figure}

\begin{figure}[htb]
\begin{center}
\includegraphics[width=15 cm]{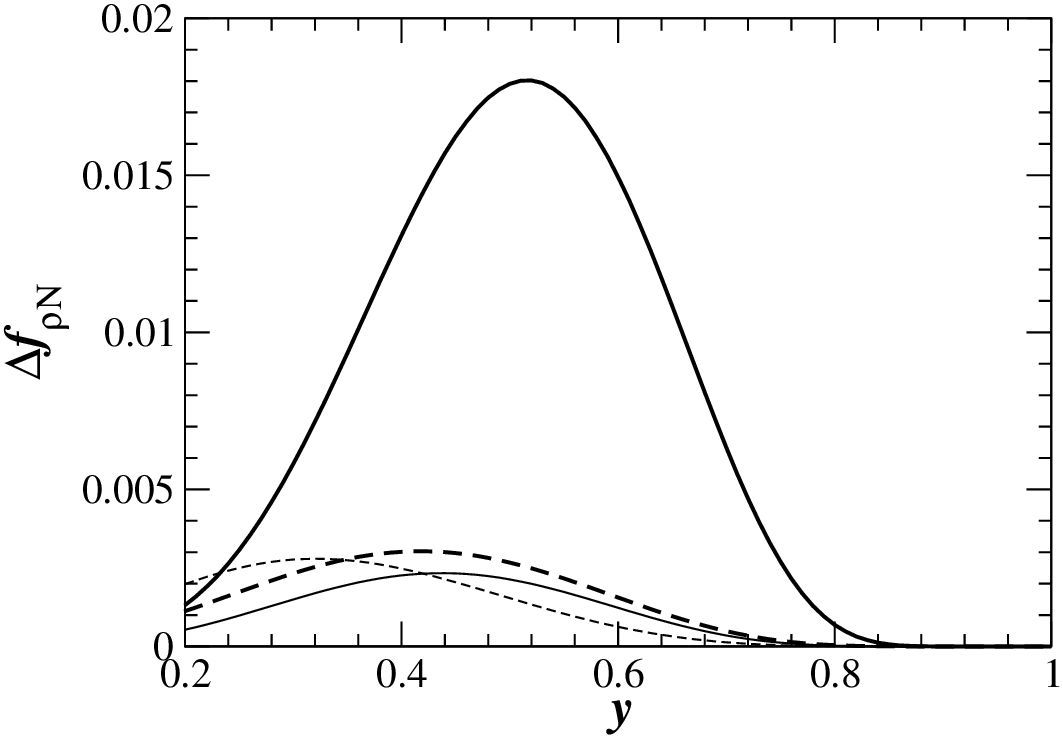}
\vskip -3.0cm
\caption{Fluctuation functions for $N \ra N \rho$ with $\rho$ being struck.
  The thick solid and dashed curves
  are for $\Delta f_{1L}$ and  $\Delta f_{2L}$, respectively. 
  The thin solid and dashed curves stand for
  $10 \, \Delta f_{1T}$ and $100 \, \Delta f_{2T}$, respectively.} 
\label{fig:DfrhoN}
\end{center}
\end{figure}

\begin{figure}[htb]
\begin{center}
\includegraphics[width=15 cm]{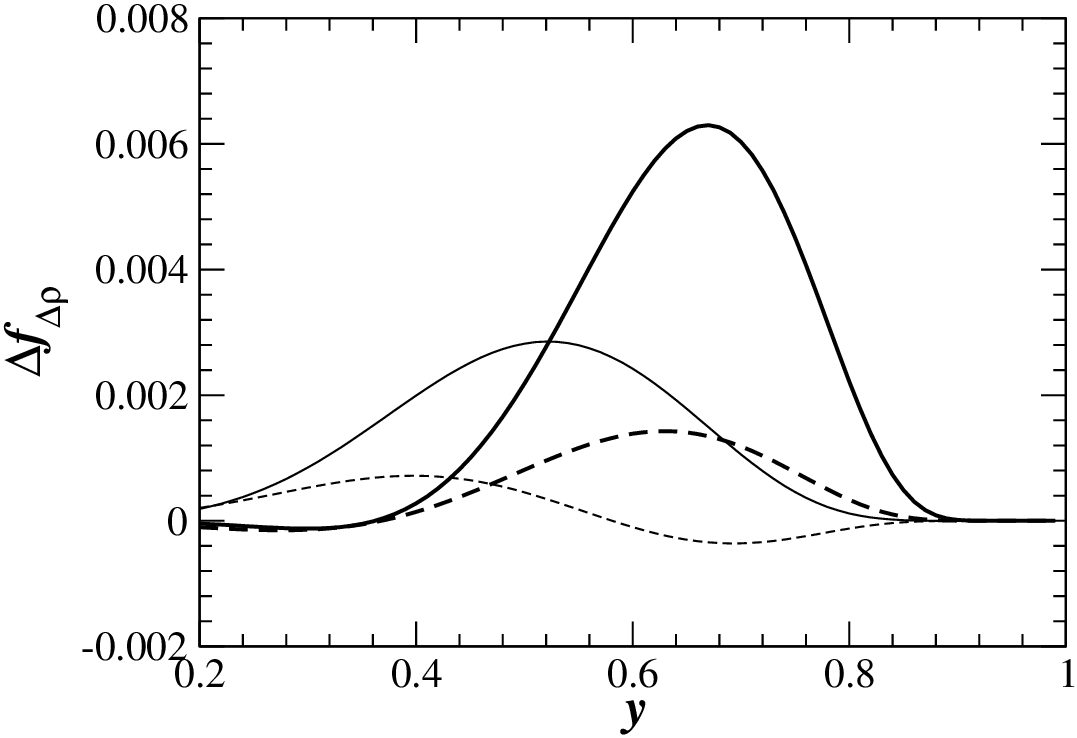}
\vskip -3.0cm
\caption{Fluctuation functions for $N \ra \Delta \rho$ with $\Delta$ being struck.
  The thick solid and dashed curves
  are for $\Delta f_{1L}$ and  $\Delta f_{2L}$, respectively. The thin solid and dashed curves stand for
  $10 \, \Delta f_{1T}$ and $100 \, \Delta f_{2T}$, respectively.
  $\Delta f_{1(2)L(T)}=\Delta f_{1(2)L(T)}^{\frac{3}{2}\frac{1}{2}}+
  3 \Delta f_{1(2)L(T)}^{\frac{3}{2}\frac{3}{2}}$.}
\label{fig:DfDrho}
\end{center}
\end{figure}

\begin{figure}[htb]
\begin{center}
\includegraphics[width=15 cm]{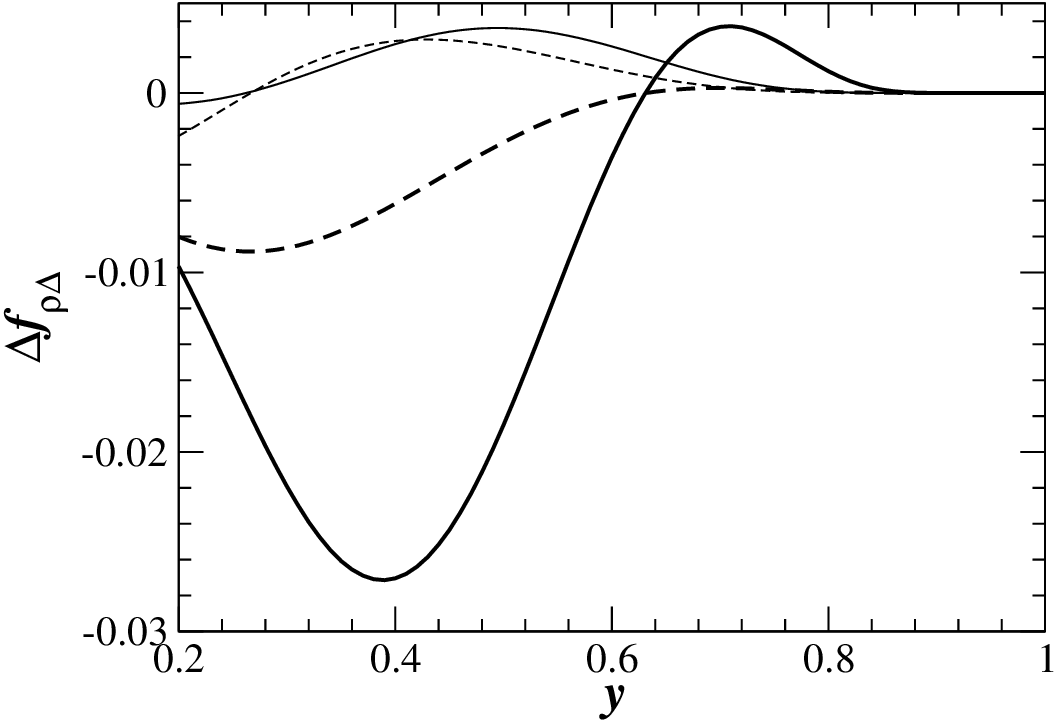}
\vskip -3.0cm
\caption{Fluctuation functions for $N \ra \Delta \rho$ with $\rho$ being struck.
  The thick solid and dashed curves
  are for $\Delta f_{1L}$ and  $\Delta f_{2L}$, respectively. The thin solid and dashed curves stand for
  $10 \, \Delta f_{1T}$ and $100 \, \Delta f_{2T}$, respectively.
  $\Delta f_{1(2)L(T)}=\Delta f_{1(2)L(T)}^{\frac{3}{2}\frac{1}{2}}+
  3 \Delta f_{1(2)L(T)}^{\frac{3}{2}\frac{3}{2}}$.}
\label{fig:DfrhoD}
\end{center}
\end{figure}

The Fock states we consider are $|N \pi \rangle$, $|N \rho \rangle$, $|\Delta \pi \rangle$, and
$|\Delta \rho \rangle$. 
The coupling constants and probabilities for each of these states in the nucleon wavefunction 
are shown in Tables~ \ref{table:couplings} and \ref{table:cutoffs}.
The effect of increasing one or both cut-offs is to increase the probability for the states 
controlled by the cut-off, and correspondingly decrease the probability of finding the 
`bare' nucleon. 
Also the probability for higher mass Fock states increases faster with the cut-off than the 
probability for lower mass states, so increasing e.g. $\Lambda_{1}$ increases the ratio of 
$|N \rho \rangle$ states to $|N \pi \rangle$ states.
The contributions from Fock states involving higher invariant mass squared
are very small, at most a few percent of the contributions from the states we consider here. 
In figures~\ref{fig:DfNpi} -- \ref{fig:DfrhoD} we show the fluctuation functions $f_{NBM}(y)$ 
for each of the Fock states.
In each of these calculations we take $x=0.2$ and $Q^2 = 2.5$ GeV$^2$, 
i.e. $\gamma^{2} = 0.056$.
In general we see that the longitudinal functions $\Delta f_{1L,2L}$ are much larger that the 
transverse functions $\Delta f_{1T,2T}$. 
This means that the contributions to $g_{1}$ of the nucleon coming from $g_{2}$ of the struck 
hadron will be small, whereas the kinematic contributions to $g_{2}$ of the nucleon from 
$g_{1}$ of the struck hadron should not be ignored as they will generally be larger than 
the contributions coming from $g_{2}$ of the Fock state hadrons.
We also observe that the fluctuation functions arising from $| \Delta \pi \rangle$ states are 
generally much larger than those of the other Fock states we consider. 
Amplitudes with the $\Delta$ having $s = 3/2$ are particularly important.  
We therefore expect that the $| \Delta \pi \rangle$ fluctuation will play a very important role in the 
MCM contributions to the spin structure functions.
Fluctuation functions involving $\rho$ mesons should not be neglected either, as these are of 
similar size to the $|N \pi \rangle $ fluctuation functions. 

\section{Polarized structure functions of bare hadrons}

To use the meson cloud model we need to know the polarized structure functions of all the 
baryons and mesons in the Fock expansion of the nucleon wavefunction.
At present the only polarized structure functions that are known experimentally are those 
of the proton and neutron (apart from the trivial case of the pseudoscalar or scalar mesons). 
It would appear unlikely that the structure functions of the $\rho$ mesons and 
the $\Delta$ baryons, which are the most important polarized cloud constituents, 
will be measured in the near future.
Our approach therefore is to estimate all the structure functions we require, including 
those of the nucleons, using the MIT bag model \cite{DeGrand75, Thomas84} and the methods 
developed by the Adelaide group \cite{CBorosT, Bag_Adelaide} and ourselves 
\cite{Signal97, FCaoS_ps2}. 
This approach gives a reasonable description of the unpolarized structure 
functions of the nucleons when compared to experimental data. 

In the bag model the dominant contributions to the parton distribution functions of
a hadron in the medium-$x$ range come from intermediate states with the lowest
number of quarks, so the intermediate states we consider contain one quark
(or anti-quark) for the mesons and two quarks for the baryons.
Following \cite{Bag_Adelaide} we can write these contributions as
\bea
q_{h,f}^{\ua \da}(x) = 
\frac{M_{h}}{(2\pi)^{3}} \sum_{m} \langle \mu | P_{f,m}| \mu \rangle 
\int d{\bf p}_{n} \frac{|\phi_{i}({\bf p}_{n})|^{2}}{|\phi_{j}({\bf 0})|^{2}} 
\delta(M_{h}(1-x) - p_{n}^{+}) |\tilde{\Psi}^{\ua \da}_{+,f}({\bf p}_{n})|^{2},
\label{mit2q}
\eea
where $M_h$ is the hadron mass,
`+' components of momenta are defined by $p^+ = p^0 + p^3$, and
${\bf p}_{n}$ is the 3-momentum of the intermediate state.
$\tilde{\Psi}$ is the Fourier transform of the MIT bag ground state wavefunction
$\Psi({\bf r})$, and $\phi_{m}({\bf p})$ is the Fourier transform of the Hill-Wheeler 
overlap function between $m$-quark bag states:
\bea
|\phi_{m}({\bf p})|^{2} = \int d{\bf R} e^{-i{\bf p \cdot R}}
\left[ \int d{\bf r} \Psi^{\dagger}({\bf r-R}) \Psi({\bf r}) \right]^{m}.
\eea
In Eq.~ (\ref{mit2q}) we take $i=1, \, j=2$ for the mesons
and $i=2, \, j=3$ for the baryons.
The matrix element $\langle \mu | P_{f,\lambda}| \mu \rangle$ appearing in  
Eq.~(\ref{mit2q}) is the matrix element of the projection operator 
$P_{f,m}$ onto the required flavour $f$ and helicity $\lambda$ for the $SU(6)$ 
spin-flavour wavefunction $| \mu \rangle$ of the hadron under consideration.

The input parameters in the bag model calculations are 
the bag radius $R$, the mass of the quark (anti-quark) $m_{q}$
for which the parton distribution is calculated, the mass of the intermediate state $m_{n}$,
and the bag scale $\mu^{2}$ -- at this scale the model is taken 
as a good approximation to the valence structure of the hadron. 
The natural scale for the model is set by the typical quark $k_{\perp}$, which is
around 0.4 GeV.
In Table~\ref{table:bag} we list the values for these parameters adopted in this work.
These values have previously been shown to give a good description of the unpolarized 
nucleon parton distributions \cite{CBorosT} and also of the meson distributions \cite{FCaoS_ps2}. 
After calculating the hadron structure functions at the bag scale $\mu^{2}$, we evolve 
them using NLO evolution \cite{Evolution} to the scale of $Q^2 = 2.5$ GeV$^2$
where the HERMES results for $g_1^p$ \cite{g1p_HERMES} and 
$g_1^n$ \cite{g1n_HERMES} are available. 
While NLO evolution from the bag scale appears stable, it would be of interest to compare 
with NNLO evolution, as it is known that LO evolution does not give very good results 
for this procedure \cite{HoytS}.
Unfortunately the necessary NNLO coefficient functions and anomalous dimensions for spin 
dependent structure functions and parton distributions have not all been calculated at 
this time.

\begin{figure}[htb]
\begin{center}
\includegraphics[width=15 cm]{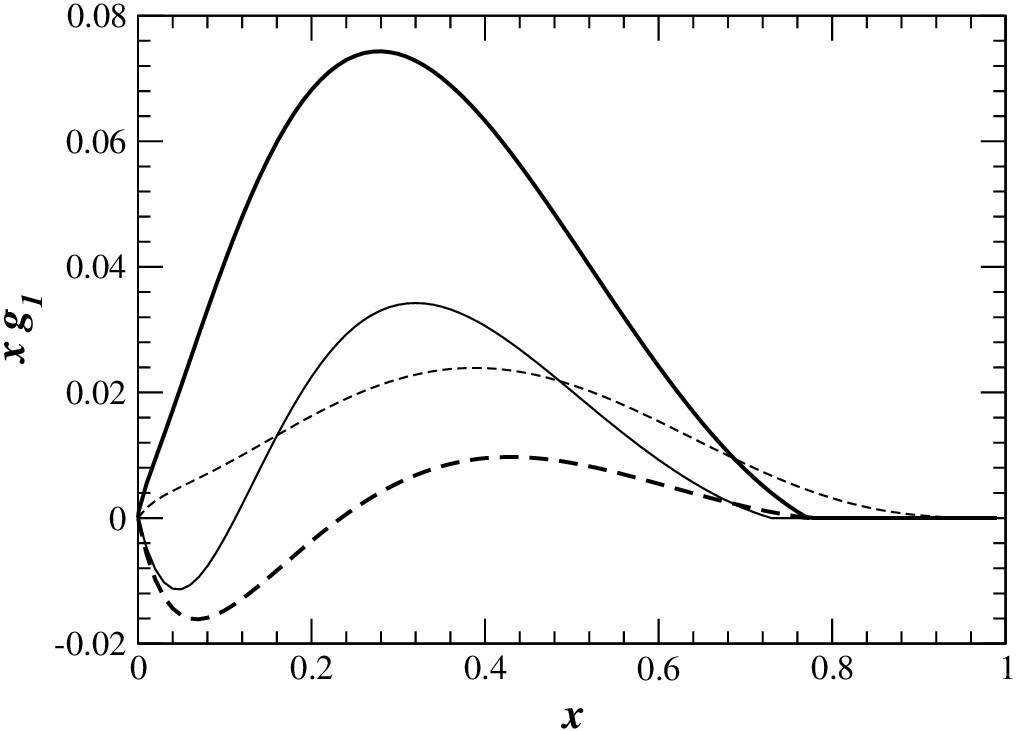}
\vskip -3.0cm
\caption{`Bare' structure functions $x g_{1}(x)$ of the nucleons, delta baryons and $\rho$ meson 
at $Q^2=2.5$ GeV$^2$. The thick solid and dashed lines are for the proton and neutron, 
respectively. The thin solid line is for the $\Delta^{+}$ baryon and the thin dashed line is 
for the $\rho$ meson.
} 
\label{fig:g1_bare}
\end{center}
\end{figure}

The calculated polarized structure functions $g_{1}^{p}(x)$ and $g_{1}^{n}(x)$ agree
reasonably well with experimental data at medium and large $x$, however they do 
not give a good description of the low $x$ data
(see the thin solid curves in figures \ref{fig:xg1p} and \ref{fig:xg1n}).
This discrepancy may well arise because our bag model calculations cannot estimate 
the polarized gluon distribution $\Delta g(x)$, which is believed to play an important role 
in the observed spin of the nucleons \cite{dgth}.
We can add a phenomenological $\Delta g(x)$,
\bea
\Delta g(x) = N_g (1-x)^\alpha.
\label{eq:Deltag}
\eea
to our calculated structure functions, where global analyses of polarized deep inelastic 
scattering data suggest $\alpha$ should be in the range of 7 -- 10 \cite{polanal}. 
In this work we use a value of $\alpha = 10$, however we have found that the calculated 
structure functions are not very sensitive to the value of $\alpha$. 
We normalise the polarized gluon distribution so that the contribution of polarized 
gluons to the first moment of $g_1^p$ (Ellis-Jaffe sum rule) and $g_1^n$ is $-0.05$, 
which gives theoretical moments that are in agreement with experiment.
The structure functions $g_1$ for the bare proton, neutron, $\Delta^+$ and $\rho$
are given in figure \ref{fig:g1_bare}.
We have not taken account of any possible topological contributions to the singlet 
axial charge $g_{A}^{(0)}$, which can also contribute to $g_{1}(x)$ at $x = 0$ \cite{Bass}. 
As our procedure gives a reasonable description of the observed $g_1^p(x)$ and $g_1^n(x)$, 
especially once MCM contributions have been added (see below), there appears little need 
to add an extra phenomenological term to the bare structure functions.

The structure functions $g_2$ for the bare hadrons are estimated
via the Wandzura-Wilczek relation \cite{WW} which is
obtained by considering only twist-2 contributions to $g_1$ and $g_2$,
\bea
g_2^{WW}(x,Q^2)=-g_1(x,Q^2)+\int^1_x\frac{dy}{y} g_1(y,Q^2)
\label{eq:g2WW}
\eea
We also note that previous bag model studies of $g_{2}(x)$ \cite{Signal97, Stratmann, Song} 
accord reasonably well with the experimental data.

\section{Numerical Results and Discussions}

\begin{figure}[htb]
\begin{center}
\vskip -2.0cm
\includegraphics[width=12 cm]{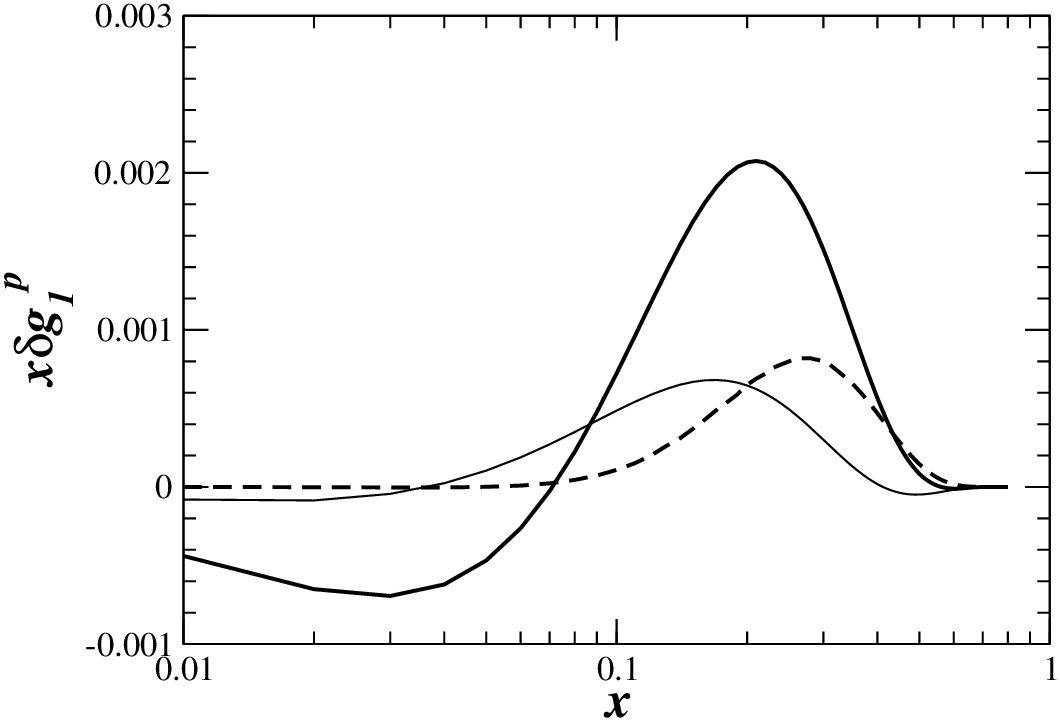}
\vskip -8.0cm
\includegraphics[width=12 cm]{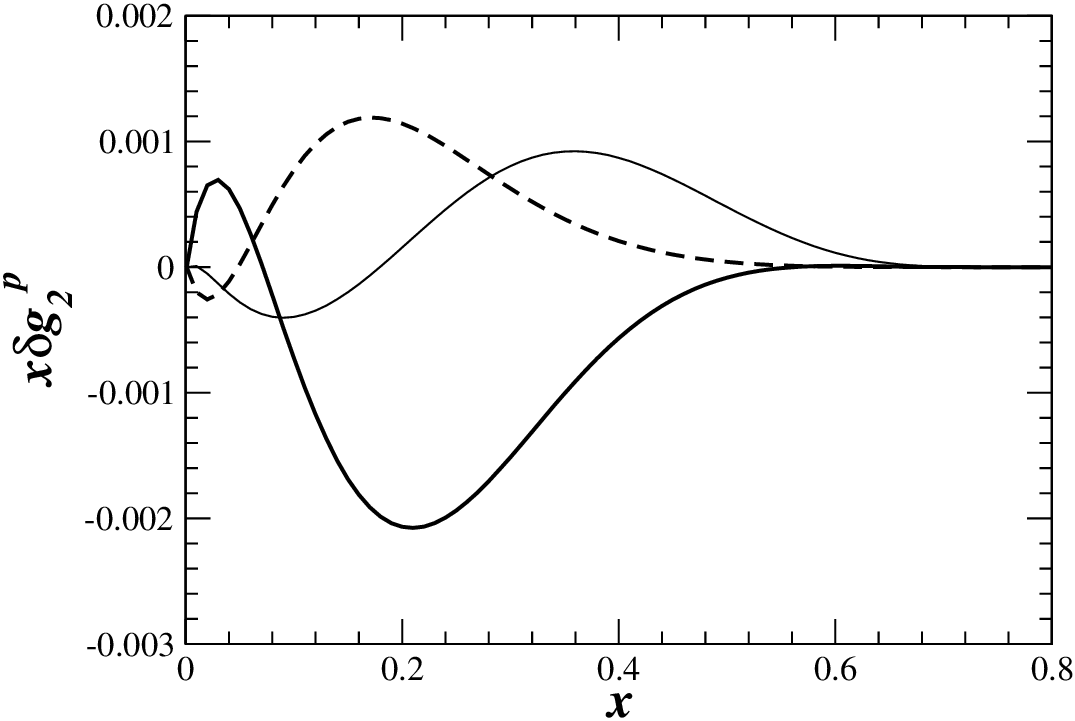}
\vskip -8.0cm
\caption{Meson cloud model contributions to $g_{1}^{p}(x)$ and $g_{2}^{p}(x)$ at $Q^2=2.5$ 
GeV$^2$. The thick lines are the total longitudinal contributions to the structure functions. 
The thick dashed lines are the total transverse contributions to the structure functions (multiplied by 
10 in the case of $g_{1T}(x)$). The thin lines show the total contribution of the $|N \pi \rangle$ Fock 
state to the structure functions (multiplied by 5 in the case of $g_{2}(x)$).
} 
\label{fig:mcmg12}
\end{center}
\end{figure}

We show our results for the MCM contributions to $g_{1}^{p}(x)$ and $g_{2}^{p}(x)$ at $Q^2 = 2.5$
GeV$^{2}$ in figure~\ref{fig:mcmg12}. 
As expected, the dominant contributions to $g_{1}^{p}$ are the longitudinal contributions of the 
form $\Delta f_{iL} \otimes g_{i}$, while the transverse contributions are fairly small at this scale. 
For $g_{2}^{p}$ the transverse contributions are similar in size to the longitudinal contributions, 
but tend to be opposite in sign, which makes the overall MCM contribution to $g_{2}^{p}$ smaller 
than for $g_{1}^{p}$. 
We also show the contribution from the $|N \pi \rangle$ portion of the MCM wavefunction. 
While important, it is clear that taking into account only this part of the wavefunction does not give 
a good indication of the total MCM contribution to the spin dependent structure functions.
The $\Delta$ baryon, especially the $s=3/2$ component, plays an important role and 
should not be ignored.

In figures~\ref{fig:xg1p} and \ref{fig:xg1n} we compare theoretical calculations
for $g_1^p$ and $g_1^n$ with recent experimental
measurements from HERMES Collaboration \cite{g1p_HERMES,g1n_HERMES}.
The calculation of $g_{1}^{n}(x)$ agrees very well with the data, but the agreement is 
less impressive in the case of $g_{1}^{p}(x)$, where the calculated structure function 
is significantly smaller than the data points in the region $x > 0.3$, and the peak of the 
calculated structure function occurs near $x = 0.3$. 
As can be seen in figure~\ref{fig:xg1p}, the fit to the experimental data is considerably improved
by including both the polarized gluon distribution and meson cloud effects.
It is known \cite{CBorosT,FSteffensHT} that the meson cloud lowers the bag model calculation for 
$g_{1}^{p}$ over the entire range of $x$ since the angular momentum of the meson cloud carries 
some of the spin of the nucleon.
However these calculations overestimate $g_{1}^{p}(x)$ in the region $x < 0.1$ and give results 
with much smaller magnitude than the experimental data for $g_{1}^{n}(x)$  in the region $x<0.2$. 
Including the polarized gluon distribution significantly improves the fit to the experimental data.
The importance of these polarized gluon contributions is more obvious in the calculation of 
the structure function $g_1^n$. 
Without these contributions our theoretical calculations are not able to reproduce the
shape of the experimental data. 
We note that the magnitude of the polarized gluon distribution we use is determined only by the 
Ellis-Jaffe sum rule, and we have not attempted to change the shape of the distribution to 
improve the agreement with the data. 
A harder polarized gluon distribution would reduce $g_{1}^{p}$ even more at large $x$.
Another factor which affects the quality of our fit to the data at large $x$ is the difficulty the 
bag model calculation of structure functions has in this region because of the non-relativistic 
projection used to form momentum eigenstates \cite{Bag_Adelaide}, which results in the 
calculated distributions being systematically smaller than the data.

The values for the Ellis-Jaffe sum rule for the proton and neutron are found to be
$0.120$ and $-0.027$ respectively, which are close to the experimental values at this scale of 
$\int_{0.021}^{0.85} g_{1}^{p}(x) dx = 0.122 \pm 0.003$ (stat.)$\pm 0.010$ (sys.) 
\cite{g1p_HERMES} and
$-0.037 \pm 0.013$ (stat.)$\pm 0.005$ (sys.)$\pm 0.006$ (extrap.) \cite{g1n_HERMES}. 
The Bjorken sum rule is found to be $0.147$ which is close to the experimental value, and 
can also be compared with the theoretical value calculated to $O(\alpha_{S}^{3})$ \cite{Larin97} 
of $0.173$. 
This is consistent with value of $g_{A}$ calculated in the bag model being 10\% smaller than 
the experimental value. 
In this work, the structure functions have been calculated by considering only the
contributions from the valence partons. 
In the MCM, the polarized antiquark distributions $\Delta \bar{u}(x)$ and $\Delta \bar{d}(x)$ are
found to be rather small \cite{FCaoS_ps2}
\footnote{While this calculation included incorrect interference terms in the calculations of 
$\Delta \bar{u}(x)$ and $\Delta \bar{d}(x)$, these terms only contributed to the spin dependent 
sea at the 10\% level.}.
However Pauli blocking effects  \cite{FCaoS_plzdsea, PauliBlocking} may be of similar size in 
the polarized distributions as in the unpolarized distributions \cite{WMelnitchoukST}, and could 
contribute $5 - 10\%$ to the observed value of the Bjorken sum rule.

Now we turn our attention to the calculations for the structure functions $g_2(x)$.
The results are presented in figures~\ref{fig:xg2p} and \ref{fig:xg2n} for the proton and neutron
respectively, along with experimental data from E99-177, E97-103 and E155 Collaborations.
The agreement between theoretical calculations and experimental measurements for the proton 
in the region $0.05<x<0.7$ is very good. 
The calculations for $g_2^n$ are consistent with the recent precision measurement at JLab for 
$x \simeq 0.2$, although experimental information on the $x$-dependence of $g_2^n(x)$ is not 
conclusive due to large error bars.
Once again we find that including the polarized gluon contribution is crucial to the calculations
for the region $x<0.2$, especially for the calculation of $g_2^n(x)$.
The cloud mesons can have a dramatic effect on the calculations for the structure functions 
$g_2^{p,n}$, especially in the region $0.1 < x <0.4$. 
For $g_2^n$ in the region of $x \sim 0.1$ the cloud contributions are
comparable in magnitude with the `bare' contributions.

The close agreement between our calculations and the agreement implies that any twist three 
portion of $g_{2}(x)$ is rather small. 
We note that experimental data from E155 \cite{A2g2_E155} is compatible, within two standard 
deviations, with there being no twist three contribution to the structure functions. 
If precision experiments at low values of $Q^{2}$ also show no firm evidence for twist three 
contributions to $g_{2}$, this will provide a new challenge for model builders, as it is expected 
that higher twist parts of the structure functions will be of similar size to leading twist contributions 
at the model scale \cite{Signal97}.
We give our results for the first few moments of $g_{2}^{p}$ and $g_{2}^{n}$, along with the 
experimental estimates of these moments in table~\ref{table:moments}. 
The disagreement between our value of the second moment of $g_{2}^{p}$ and that of E155
is largely due to the data point at $x = 0.78$ which gets a large weighting in the calculation 
of the moment.

\section{Summary}

We have used the meson cloud model to calculate the spin dependent structure functions 
$g_{1}(x)$ and $g_{2}(x)$ of the proton and neutron. 
An important part of this calculation is the use of `bare' structure functions of the hadrons 
in the model calculated in the MIT bag model, with the addition of an extra polarized gluon 
term, which gives good agreement with the observed value of the lowest moment of $g_{1}^{p}$ 
(Ellis-Jaffe sum rule).

We included in our calculations the full effects of kinematic terms that arise at finite $Q^{2}$ 
because the spin vector of the struck hadron is not parallel with the spin vector of the initial nucleon. 
This leads to three or more additional contributions to each spin dependent structure function for 
each hadron species included in the model nucleon wavefunction. 
While these contributions vanish in the Bjorken limit, they can make a substantial proportion of the 
observed structure functions at finite $Q^{2}$, and are particularly important for describing the 
neutron structure functions. 
We note that these contributions have the same form as expected for target mass corrections 
\cite{Roberts} and should not be confused with genuine higher twist contributions to the 
structure functions, which arise from new operators involving quark-gluon correlators 
\cite{Signal97, Ji93}.
As the quality of data on neutron structure functions improves, it will be interesting to compare 
the behaviour of the structure functions as a function of $Q^{2}$ with that predicted by the MCM.
We have not done this here, as most of the data is at fairly low $x$, and the $Q^{2}$ variation in 
this kinematic region is quite small. 

We have considered the effects of possible interference between intermediate states containing 
different hadrons, which can contribute to the spin dependent parts of the DIS cross section. 
Our analysis shows that for the most part these terms cannot affect the observed structure 
functions or parton distributions.
It is possible that states involving higher spins {\it e.g.} $| \Delta \rho \rangle$ can give 
interference contributions. 
The difficulties in calculating the dynamics of the relevant vertices are formidable, however these 
terms are suppressed in the MCM owing to their high mass, so we have ignored these contributions 
in this work. 

Our calculations of the spin dependent structure functions show good agreement with the 
experimental data. 
We see significant corrections to the structure functions calculated using the bag model arising 
both from the inclusion of a polarized gluon distribution and from the cloud contributions. 
In both cases these improve the agreement  with the experimental data. 
Our calculations of $g_{2}(x)$ includes only the Wandzura-Wilczek term, which gives the twist 
two portion of the structure function. 

There is a further spin dependent structure function of the nucleon, which we have not discussed 
in this paper.
This is the transversity distribution $h_{1}(x)$, which measures the distribution of transversely 
polarized quarks in a transversely polarized nucleon. 
In the non-relativistic limit $h_{1}(x) = g_{1}(x)$, so a measurement of $h_{1}$ can tells us about 
the importance of relativistic effects in the quark wavefunction.
We will be extending our calculations to the transversity distribution of the proton and neutron, 
where we expect that meson cloud model effects will affect significantly the observed 
structure functions.
 
\section*{Acknowledgments}
We thank Tony Thomas for helpful comments and encouragement. This 
work was partially supported by 
the Marsden Fund of the Royal Society of New Zealand. 
F.B. is supported by a Massey University Post-doctoral Fellowship.
A.S. acknowledges the hospitality and support of the Institute for Particle Physics
Phenomenology, Durham University, where portions of this work were done.

\section*{Appendix A Baryon and Meson momentum distributions}

We reproduce here the results for the general form of the baryon and meson momentum 
distributions $f^{\lambda}_{1,2 L,T}(y)$ as given by Kumano and Miyama \cite{KM} for 
spin $1$ mesons. 
As we have noted above, these results also hold for spin $1/2$ baryons and can be generalised 
to spin 3/2 baryons. 
Firstly we note that the portion of the integrand of equation (\ref{eq:Bcoeffs}) that 
depends on $J_{NMB}$, the NMB vertex times the propagator of the struck hadron, may 
be written as a sum of an unpolarized, a longitudinal and a transverse part:
\bea
\frac{|\vec{p}_{N}|}{(2 \pi)^{3}} y \frac{\partial y^{\prime}}{\partial y} 
\sum_{\lambda^{\prime}} |J_{NBM}|^{2} = C_{0}^\lambda + \lambda_N C_{L}^{\lambda} + \tau_{N} \cos \phi  \, 
C_{T}^{\lambda}
\eea
where $\phi$ is the angle between $\vec{k}_{\perp}$ and $\vec{s}_{N}^{\top}$, 
$\vec{k}_{\perp} \cdot \vec{s}_{N}^{\top} = |\vec{k}_{\perp}| \tau_{N} \cos \phi$,
 and $\lambda$ labels the helicity of the struck hadron.
We find that the unpolarized part is such that $C_0^{-\lambda}=C_0^\lambda $, so any 
contribution arising from this part will cancel itself when we compute $\Delta f$ using
equation (\ref{eq:deltaf}). Therefore we will ignore any contributions from $C_0^\lambda$ 
in the following.
Performing the integration over $\phi$ then gives equation (\ref{eq:totconv}) with the 
hadron momentum distributions given by
\bea
f_{1 L,T}^{\lambda}(y) & = & \int_{0}^{(k_{\perp}^{2})_{m}} d \vec{k}_{\perp}^{2} \, 
r_{1 L,T}^{\lambda}(\vec{k}_{\perp}^{2}, m) \\
f_{2 L,T}^{\lambda}(y) & = & \int_{0}^{(k_{\perp}^{2})_{m}} d \vec{k}_{\perp}^{2} \,
r_{2 L,T}^{\lambda}(\vec{k}_{\perp}^{2}, m)
\eea
where $m$ is the mass of the struck hadron and the integrands $r_{1,2 i}^{\lambda}$ are 
given by
\bea
r_{1 L}^{\lambda}(\vec{k}_{\perp}^{2}, m) & = &  2 \pi C_{L}^{\lambda} 
\left[ 1 + \frac{k_{\perp}^{2}}{y y^{\prime} m_{N}^{2}}(\sqrt{1 + \gamma^{2}} - 1) \right] \\
r_{1 T}^{\lambda}(\vec{k}_{\perp}^{2}, m) & = & - \gamma^{2} \pi C_{T}^{\lambda} 
\frac{k_{\perp}}{y m_{N}} \\
r_{2 L}^{\lambda}(\vec{k}_{\perp}^{2}, m) & = & - \gamma^{2} \pi C_{L}^{\lambda} 
\frac{m^{2}}{y^{2} m_{N}^{2}} \\
r_{2 T}^{\lambda}(\vec{k}_{\perp}^{2}, m) & = &  \gamma^{2} \pi C_{T}^{\lambda} 
\frac{k_{\perp}m^{2}}{y^{2} y^{\prime} m_{N}^{3}}(\sqrt{1 + \gamma^{2}} - 1).
\eea
We note that we have changed the signs of $f_{1 T}$ and $f_{2 L}$ from those of \cite{KM}, 
as this is more consistent with the notation we use below for the spin $3/2$ baryon momentum 
distributions 
In the Bjorken limit only $f_{1 L}$ remains non-zero. 
By combining the longitudinal $\lambda_{N} = 1\, (\tau_{N}=0)$ with the transverse 
$\lambda_{N} = 0\, (\tau_{N}=1)$ amplitude in equation (\ref{eq:totconv}), and defining the 
functions
\bea
\Delta f_{1,2 L,T}^{j}(y) = f_{1,2 L,T}^{\lambda = +j}(y) - f_{1,2 L,T}^{\lambda = -j}(y)
\eea
with $j$ the spin of the struck hadron, we can obtain the separate contributions to the nucleon 
$g_{1}$ and $g_{2}$ structure functions:
\bea
\delta g_{1}(x, Q^{2}) & = & \frac{1}{1+\gamma^{2}} \int_{x}^{1} \frac{dy}{y} \sum_{i = 1,2} 
[\Delta f_{i L}^{j}(y) - \Delta f_{i T}^{j}(y)] g_{i}^{j} \left( \frac{x}{y}, Q^{2} \right) \\
\delta g_{2}(x, Q^{2}) & = & -\frac{1}{1+\gamma^{2}} \int_{x}^{1} \frac{dy}{y} \sum_{i = 1,2} 
\left[\Delta f_{i L}^{j}(y) + \frac{\Delta f_{i T}^{j}(y)}{\gamma^{2}} \right] g_{i}^{j} \left( \frac{x}{y}, Q^{2} \right).
\eea

The momentum distributions for spin $3/2$ baryons follow a similar pattern to thse above.
In this case the distributions also have to be labelled by $h = \frac{1}{2},\,\frac{3}{2}$ 
with $\lambda = \pm h$. 
We obtain
\bea
f_{1 L,T}^{\frac{3}{2} \lambda}(y) & = & \int_{0}^{(k_{\perp}^{2})_{m}} d \vec{k}_{\perp}^{2} \, 
\left[ \frac{14 - 2\omega^{2}}{15} r_{1 L,T}^{\lambda}(\vec{k}_{\perp}^{2}, m) + 
\frac{4}{15} r_{2 L,T}^{\lambda}(\vec{k}_{\perp}^{2}, m)\right] \\
f_{2 L,T}^{\frac{3}{2} \lambda}(y) & = & \int_{0}^{(k_{\perp}^{2})_{m}} d \vec{k}_{\perp}^{2} \,
\frac{18 - 2\omega^{2}}{15} r_{2 L,T}^{\lambda}(\vec{k}_{\perp}^{2}, m) \\
f_{1 L,T}^{\frac{1}{2} \lambda}(y) & = & \int_{0}^{(k_{\perp}^{2})_{m}} d \vec{k}_{\perp}^{2} \, 
\left[ \frac{6 + 2\omega^{2}}{5} r_{1 L,T}^{\lambda}(\vec{k}_{\perp}^{2}, m) - 
\frac{4}{5} r_{2 L,T}^{\lambda}(\vec{k}_{\perp}^{2}, m)\right] \\
f_{2 L,T}^{\frac{1}{2} \lambda}(y) & = & \int_{0}^{(k_{\perp}^{2})_{m}} d \vec{k}_{\perp}^{2} \,
\frac{2 + 2\omega^{2}}{5} r_{2 L,T}^{\lambda}(\vec{k}_{\perp}^{2}, m).
\eea
Now defining 
\bea
\Delta f_{1,2 L,T}^{\frac{3}{2} l}(y) = l [f_{1,2 L,T}^{l \lambda = +l}(y) - 
f_{1,2 L,T}^{l \lambda = -l}(y)],
\eea
the contributions to the structure functions become
\bea
\delta g_{1}(x, Q^{2}) & = & \frac{1}{1+\gamma^{2}} \int_{x}^{1} \frac{dy}{y} 
\sum_{i = 1,2} \sum_{h = \frac{1}{2},\frac{3}{2}} 
[\Delta f_{i L}^{\frac{3}{2} h}(y) - \Delta f_{i T}^{\frac{3}{2} h}(y)] 
g_{i}^{\frac{3}{2} h} \left( \frac{x}{y}, Q^{2} \right) \\
\delta g_{2}(x, Q^{2}) & = & -\frac{1}{1+\gamma^{2}} \int_{x}^{1} \frac{dy}{y} 
\sum_{i = 1,2} \sum_{h = \frac{1}{2},\frac{3}{2}}
\left[\Delta f_{i L}^{\frac{3}{2} h}(y) + \frac{\Delta f_{i T}^{\frac{3}{2} h}(y)}{\gamma^{2}} \right] 
g_{i}^{\frac{3}{2} h} \left( \frac{x}{y}, Q^{2} \right).
\eea

\section*{Appendix B Calculation of $N-\Delta$ Interference terms}

As an example of the general technique for calculating the vertex functions in the MCM, and 
more specifically how to calculate interference terms, we present the calculation of the terms 
for interference between $|N \pi \rangle $  and $|\Delta \pi \rangle $ states, where the pion is
the spectator.
We start from the interaction Lagrangians for the two vertices \cite{HHoltmannSS, SpethT, RMachleidtHE}
\bea
{\cal L}_{1} & =  & i g_{NN \pi} \bar{\psi} \gamma_{5} \pi \psi \\
{\cal L}_{2} & =  & f_{N \Delta \pi} \bar{\psi} \partial_{\mu} \pi U^{\mu} + \mbox{h.c.}
\eea
where $U^{\mu}(p,s)$ is the Rarita-Schwinger spinor for the spin $3/2$ field. 
These give the required vertices in the numerators of $J_{NN \pi}$ and $J_{N \Delta \pi}$, 
whereas the denominators will be given by the propagators of the nucleon and $\Delta$ 
respectively.
Standard techniques then give us that ${ \cal R}[J_{N \Delta \pi} J_{N N \pi}^{*}]$ is 
proportional to the trace of
\bea
\frac{1}{2}[ u(p, s) \bar{u}(p,s) ]( \gamma_{5} E^{\mu}(k_{1}, s_{1}) \bar{u}(k_{2}, s_{2}) - 
u(k_{2}, s_{2})\bar{E}^{\mu}(k_{1}, s_{1}) \gamma_{5}  ) p^{\prime}_{\mu} 
\eea
where $E^{\mu}(k_{1}, s_{1})$ is the positive energy spin $3/2$ spinor, which is a linear combination 
of Dirac spinors of positive and negative helicity with polarization vectors $\epsilon^{\mu}$ for 
longitudinal and left or right circular polarization in the moving frame.
We note that the nucleon and $\Delta$ in the intermediate states do not have identical momentum 
or spin 4-vectors, which may have been previously overlooked in earlier calculations.
This fact makes the interference calculations much more difficult than the usual non-interference 
terms as $E^{\mu}(k_{1}, s_{1}) \bar{u}(k_{2}, s_{2})$ cannot be written as a propagator, but requires 
careful evaluation.
We also note that our calculation adds together two contributions depending on whether the 
initial MCM state is $|N \pi \rangle $  or $|\Delta \pi \rangle $. 
As these two are Hermitian conjugates the final result should be real, which acts as a check 
that we have correctly accounted for all phase factors.

It is easiest to do the calculation in two parts, corresponding to the $\lambda = \pm 1/2$ helicities 
of the intermediate state hadrons.
For the coefficients $C_{0,L,T}^{\lambda}$ of equation (\ref{eq:ccoeffs}) we obtain
\bea
C_{0}^{+}(\phi) & = &  C_{0}^{-}(\phi) = \frac{k_{\perp}^{2} (m_{N}+m_{\Delta})}
{4 \sqrt{6} m_{N} \sqrt{m_{N} m_{\Delta}}y^{\prime 2}} \cos \phi \\
C_{L}^{+}(\phi) & =  & - C_{L}^{- }(\phi) = \frac{k_{\perp}^{2} (m_{N}(1-2 y^{\prime}) - m_{\Delta})}
{4 \sqrt{6} m_{N} \sqrt{m_{N} m_{\Delta}}y^{\prime 2}} \cos \phi 
\eea
and
\bea
C_{T}^{+}(\phi) & = & - C_{T}^{-}(\phi) \nonumber \\
& = & \frac{(y^{\prime}-1)  \left(y^{\prime} m_N + m_{\Delta}\right)k_{\perp}}
{4 \sqrt{6}y^{\prime 2} \sqrt{m_N m_{\Delta }}} - 
\frac{  k_{\perp}^3}{4 \sqrt{6} y^{\prime 2} m_{N} \sqrt{m_N m_{\Delta}}} \cos (2 \phi ) + \nonumber \\  
&&\frac{ \left(m_N +m_{\Delta }\right) \left(k_{\perp}^2+y^{\prime 2} m_{\pi
   }^2-(y^{\prime}-1)^2 m_{\Delta }^2\right) k_{\perp}}{4 \sqrt{6} (y^{\prime}-1) y^{\prime 2}
   \left(m_N m_{\Delta }\right)^{3/2}}  \sin(\phi ).
\eea

These coefficients now must be multiplied by coefficients 
$A_{1,2}^{N \Delta}, \: \tilde{A}_{1,2}^{N \Delta}$ from equation 
(\ref{eq:Acoeffsinf}), all of which have the structure
\bea
A_{i}^{{N \Delta}} = a_{i}^{L} \lambda_{N} +  a_{i}^{T} \tau_{N} \cos \phi
\eea
and similarly for $\tilde{A}_{1,2}^{N \Delta}$,
where $\lambda_{N}$ and $\tau_{N}$ refer to the polarization of the parent nucleon. 
The functions $r_{1,2 L,T}^{\lambda}$ (and their analogous $\tilde{r}_{1,2 L,T}^{\lambda}$) 
are then given by
\bea
r_{i \, L}^{\lambda} &=& \int_{0}^{2\pi} d\phi \,a_{i}^{L} C_{L}^{\lambda} \\
r_{i \, T}^{\lambda} &=& \int_{0}^{2\pi} d\phi \,a_{i}^{T} C_{T}^{\lambda} \cos \phi
\eea
while any contribution to the unpolarized cross section will be proportional to 
\bea
\int_{0}^{2\pi} d\phi [C_{0}^{+} + C_{0}^{-}].
\eea
By inspection all these angular integrals are zero.
Thus the interference between  $|N \pi \rangle $  and $|\Delta \pi \rangle $ intermediate states 
makes no contribution to the observed structure functions. 
Similar arguments hold for the case of interference between $|N \pi \rangle $ and $|N \rho \rangle $ 
intermediate states.

\newpage
%\vskip 1cm
\section*{Figure Captions}
\begin{figure}[htb]
\begin{center}
\includegraphics[width=15 cm]{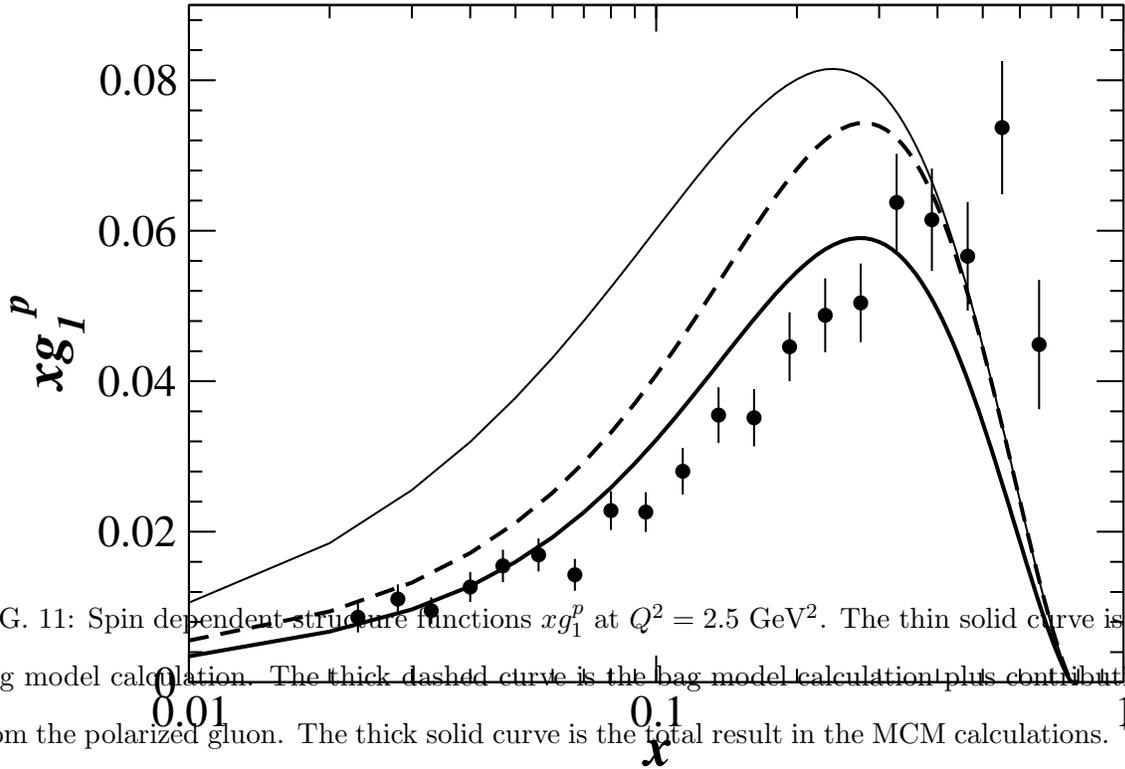}
\vskip -3.0cm
\caption{Spin dependent structure functions $x g_1^p$ at $Q^2=2.5$ GeV$^2$.
  The thin solid curve is the bag model calculation.
  The thick dashed curve is the bag model calculation plus contributions from the polarized gluon.
  The thick solid curve is the total result in the MCM calculations. 
  The HERMES data are taken from \cite{g1p_HERMES}.} 
\label{fig:xg1p}
\end{center}
\end{figure}

\begin{figure}[htb]
\begin{center}
\includegraphics[width=15 cm]{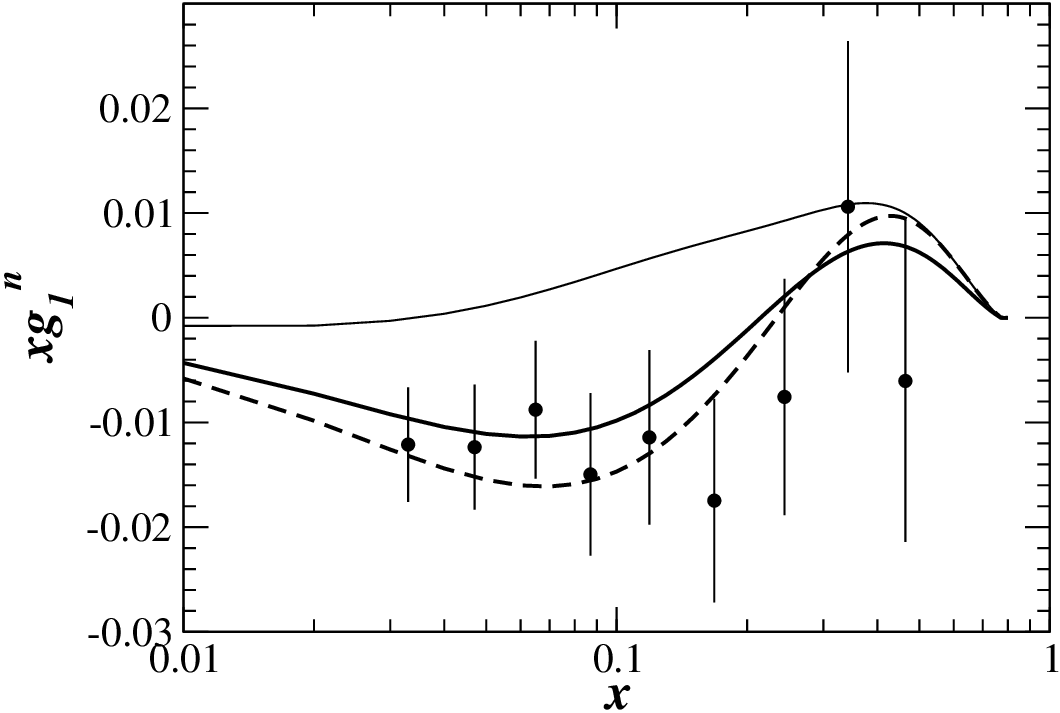}
\vskip -3.0cm
\caption{As in figure \ref{fig:xg1p} but for $x g_1^n$.
  The HERMES data are taken from \cite{g1n_HERMES}. }
\label{fig:xg1n}
\end{center}
\end{figure}

\begin{figure}[htb]
\begin{center}
\includegraphics[width=15 cm]{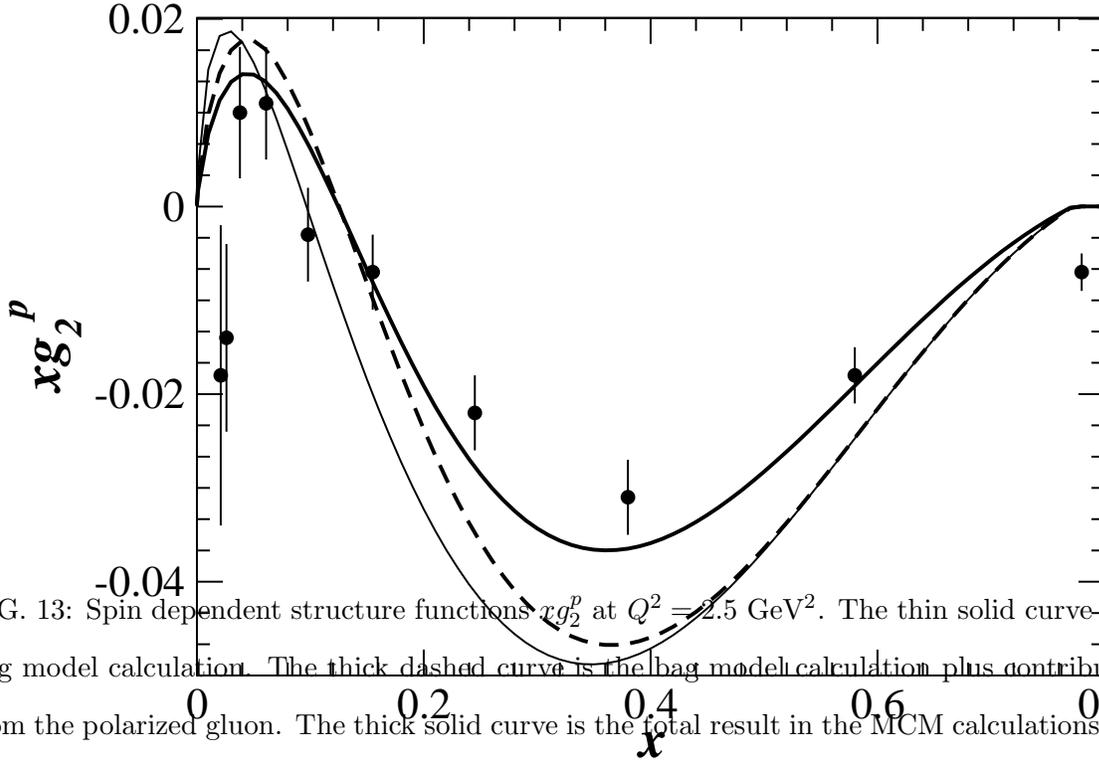}
\vskip -3.0cm
\caption{Spin dependent structure functions $x g_2^p$ at $Q^2=2.5$ GeV$^2$.
  The thin solid curve is the bag model calculation.
  The thick dashed curve is the bag model calculation plus contributions from the polarized gluon.
  The thick solid curve is the total result in the MCM calculations. 
  The data are taken from SLAC-E155 \cite{A2g2_E155} and
  $0.8 \, {\rm GeV}^2 <Q^2< 8.2 \, {\rm GeV}^2$.}
\label{fig:xg2p}
\end{center}
\end{figure}

\begin{figure}[htb]
\begin{center}
\includegraphics[width=15 cm]{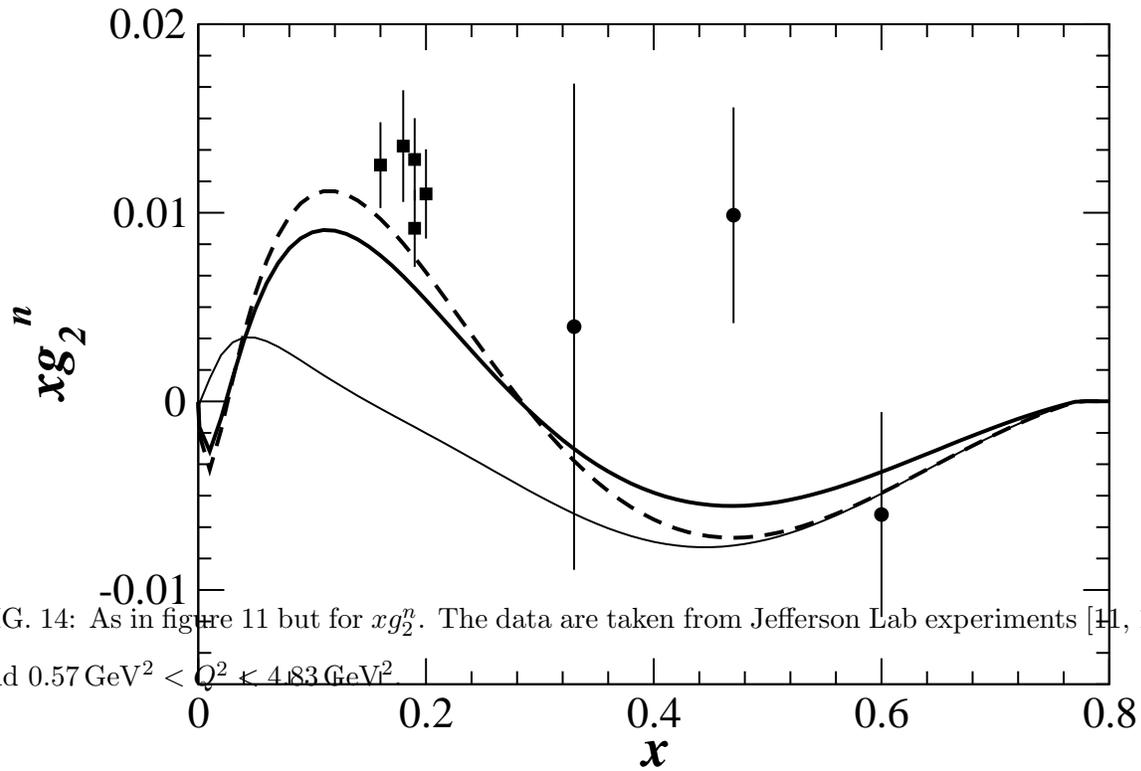}
\vskip -3.0cm
\caption{As in figure \ref{fig:xg1p} but for $x g_2^n$.
The data are taken from Jefferson Lab experiments \cite{g2HallA,g2HallA_Kramer} and
$0.57 \, {\rm GeV}^2 <Q^2< 4.83 \, {\rm GeV}^2$.}
\label{fig:xg2n}
\end{center}
\end{figure}

\begin{description}
\item
{Fig.~1.}
Contributions to the physical nucleon structure function. The photon may be scattered from 
(a) the virtual meson or (b) the virtual baryon.
\item
{Fig.~2.}
Interference between $\pi$ and $\rho$ mesons.
\item
{Fig.~3.}
Fluctuation functions for $N \ra N \pi$ with $N$ being struck.
The thick solid and dashed curves are for $\Delta f_{1L}$ and  $\Delta f_{2L}$, respectively. 
The thin solid and dashed curves stand for $10 \, \Delta f_{1T}$ and 
$100 \, \Delta f_{2T}$, respectively.
\item
{Fig.~4.}
Fluctuation functions for $N \ra \Delta \pi$ with $\Delta$ being struck.
The thick solid and dashed curves are for $\Delta f_{1L}$ and  $\Delta f_{2L}$, respectively. 
The thin solid and dashed curves stand for $10 \, \Delta f_{1T}$ and 
$100 \, \Delta f_{2T}$, respectively.
$\Delta f_{1(2)L(T)} = \Delta f_{1(2)L(T)}^{\frac{3}{2}\frac{1}{2}} + 
3 \Delta f_{1(2)L(T)}^{\frac{3}{2}\frac{3}{2}}$.
\item
{Fig.~5.}
Fluctuation functions for $N \ra N \rho$ with $N$ being struck.
The thick solid and dashed curves are for $\Delta f_{1L}$ and  $\Delta f_{2L}$, respectively. 
The thin solid and dashed curves stand for $10 \, \Delta f_{1T}$ and 
$100 \, \Delta f_{2T}$, respectively.
\item
{Fig.~6.}
Fluctuation functions for $N \ra N \rho$ with $\rho$ being struck.
The thick solid and dashed curves are for $\Delta f_{1L}$ and  $\Delta f_{2L}$, respectively. 
The thin solid and dashed curves stand for $10 \, \Delta f_{1T}$ and 
$100 \, \Delta f_{2T}$, respectively.
\item
{Fig.~7.}
Fluctuation functions for $N \ra \Delta \rho$ with $\Delta$ being struck.
The thick solid and dashed curves are for $\Delta f_{1L}$ and  $\Delta f_{2L}$, respectively. 
The thin solid and dashed curves stand for
$10 \, \Delta f_{1T}$ and $100 \, \Delta f_{2T}$, respectively.
\item
{Fig.~8.}
Fluctuation functions for $N \ra \Delta \rho$ with $\rho$ being struck.
The thick solid and dashed curves are for $\Delta f_{1L}$ and  $\Delta f_{2L}$, respectively. 
The thin solid and dashed curves stand for
$10 \, \Delta f_{1T}$ and $100 \, \Delta f_{2T}$, respectively.
\item
{Fig.~9.}
`Bare' structure functions $x g_{1}(x)$ of the nucleons, delta baryons and $\rho$ meson 
at $Q^2=2.5$ GeV$^2$. The thick solid and dashed lines are for the proton and neutron, 
respectively. The thin solid line is for the $\Delta^{+}$ baryon and the thin dashed line is 
for the $\rho$ meson.
\item
{Fig.~10.}
Meson cloud model contributions to $g_{1}^{p}(x)$ and $g_{2}^{p}(x)$ at $Q^2=2.5$ 
GeV$^2$. The thick lines are the total longitudinal contributions to the structure functions. 
The thick dashed lines are the total transverse contributions to the structure functions (multiplied by 
10 in the case of $g_{1T}(x)$). The thin lines show the total contribution of the $|N \pi \rangle$ Fock 
state to the structure functions (multiplied by 5 in the case of $g_{2}(x)$).
\item
{Fig.~11.}
Spin dependent structure functions $x g_1^p$ at $Q^2=2.5$ GeV$^2$.
The thin solid curves are the bare bag model calculations.
The thick dashed curves are the bag model calculations plus contributions from the polarized gluon.
The thick solid curves are the total results in the MCM calculations. 
The HERMES data are taken from \cite{g1p_HERMES}.
\item
{Fig.~12.}
As in figure \ref{fig:xg1p} but for $x g_1^n$.
The HERMES data are taken from \cite{g1n_HERMES}.
\item
{Fig.~13.}
Spin dependent structure functions $x g_2^p$ at $Q^2=2.5$ GeV$^2$.
The thin solid curves are the bare bag model calculations.
The thick dashed curves are the bag model calculations plus contributions from the polarized gluon.
The thick solid curves are the total results in the MCM calculations. 
The data are taken from SLAC-E155 \cite{A2g2_E155} and $0.8 \, {\rm GeV}^2 <Q^2< 8.2 \, {\rm GeV}^2$.
\item
{Fig.~14.}
As in figure \ref{fig:xg1p} but for $x g_2^n$.
The data are taken from Jefferson Lab experiments \cite{g2HallA,g2HallA_Kramer} and
$0.57 \, {\rm GeV}^2 <Q^2< 4.83 \, {\rm GeV}^2$.

\end{description}
%%%%%%%%%%%%%%%%%%%%%%%%%%%%%%%%%%%%%%%%
\newpage
\begin{table}
\vskip 0.5cm
\begin{tabular}{|c|c|c|c|c|}\hline
$\frac{g_{NN\pi}^{2}}{4\pi}$ & $\frac{f_{N\Delta \pi}^{2}}{4\pi^{2}}$ & 
$\frac{g_{NN\rho}^{2}}{4\pi}$ & $f_{NN\rho}$ & $\frac{f_{N\Delta \rho}^{2}}{4\pi}$ \\ \hline     
$13.6 $ & $12.3$ GeV$^{-2}$ & $0.84$ & $6.1 g_{NN\rho}$ & $20.45$ GeV$^{-2}$  \\ \hline
\end{tabular}
\caption{Strong coupling constants used in this work.}
 \label{table:couplings}
\end{table}

%\newpage
\begin{table}
\vskip 0.5cm
\begin{tabular}{|c|c|c|c|c|c|c|}\hline
$\Lambda_{1}$(GeV) & $\Lambda_{2}$(GeV) & 
$P_{N\pi}$ & $P_{\Delta \pi}$ & $P_{N\rho}$ & $P_{\Delta\rho}$ & $Z$ \\ \hline     
$0.8 $ & $1.0$ & $0.132$ & $0.118$ & $0.015$ & $0.025$ & $0.775$ \\ \hline
$1.0 $ & $1.0$ & $0.252$ & $0.118$ & $0.106$ & $0.025$ & $0.666$ \\ \hline
\end{tabular}
\caption{Meson Cloud Model cut-off parameters and probabilities. 
$Z$ is the wave funtion renormalization $Z = (1 + \sum_{BM} P_{BM})^{-1}$.
In this paper we have used $\Lambda_{1} = 0.8$ GeV and $\Lambda_{2} = 1.0$ GeV. 
We also display the probabilities obtained using the cut-offs of the J\"{u}lich 
group \cite{HHoltmannSS}.}
 \label{table:cutoffs}
\end{table}

%\newpage
\begin{table}
\vskip 0.5cm
\begin{tabular}{|c|c|c|c|c|}\hline
           & $R$(fm) & $m_q$(MeV) & $m_n$(MeV) & $\mu^2$(GeV$^2$) \\ \hline     
$N $   & $0.8$    &  $0$                 & $800$             & $0.23$ \\ \hline
$\rho$ & $0.7$   &  $0$                 & $425$             & $0.23$ \\ \hline
\end{tabular}
\caption{Input parameters for the bag model calculation of 'bare' structure functions.}
\label{table:bag}
\end{table}

%\newpage
\begin{table}
\vskip 0.5cm
\begin{tabular}{|c|c|c|c|}\hline
& ${\cal M}_{2}[g_{2}^{p}] \times 10^{3}$ & ${\cal M}_{4}[g_{2}^{p}] \times 10^{3}$ & 
${\cal M}_{6}[g_{2}^{p}] \times 10^{3}$ \\ \hline
This work &  -5.13 &   -1.14  &  -0.33  \\ 
Experiment \cite{A2g2_E155} & $-7.2 \pm 0.5 \pm 0.3 $ &  &   \\ \hline
& ${\cal M}_{2}[g_{2}^{n}] \times 10^{3}$ & ${\cal M}_{4}[g_{2}^{n}] \times 10^{3}$ & 
${\cal M}_{6}[g_{2}^{n}] \times 10^{3}$ \\ \hline
This work &   -0.564  & -0.203  & -0.067  \\ 
Experiment \cite{E143} &  $3.3 \pm 6.5$  &    &   \\ \hline
\end{tabular}
\caption{Comparison of our calculations with experiment for the moments of the $g_{2}$ structure 
functions, where ${\cal M}_{n}[g_{2}] = \int_{0}^{1} x^{n} g_{2}(x) dx$.}
\label{table:moments}
\end{table}

\FloatBarrier

%%%%%%%%%%%%%%%%%%%%%%%%%%%%%%%%%%%%%%%%%%%%%%%%
\newpage
%%%%%%%%%%%%%%%%%%%%%%%%%%%%%%%%%%%%%%%%%%%%%%%%

\newpage
%%%%%%%%%%%%%%%%%%%%%%%%%%%%%%%%%%%%%%%%%%%%%%%%

\newpage
%%%%%%%%%%%%%%%%%%%%%%%%%%%%%%%%%%%%%%%%%%%%%%%%

\newpage
%%%%%%%%%%%%%%%%%%%%%%%%%%%%%%%%%%%%%%%%%%%%%%%%

\newpage
%%%%%%%%%%%%%%%%%%%%%%%%%%%%%%%%%%%%%%%%%%%%%%%%

\newpage
%%%%%%%%%%%%%%%%%%%%%%%%%%%%%%%%%%%%%%%%%%%%%%%%

\newpage
%%%%%%%%%%%%%%%%%%%%%%%%%%%%%%%%%%%%%%%%%%%%%%%%
\newpage
%%%%%%%%%%%%%%%%%%%%%%%%%%%%%%%%%%%%%%%%%%%%%%%%

\newpage
%%%%%%%%%%%%%%%%%%%%%%%%%%%%%%%%%%%%%%%%%%%%%%%%

\newpage
%%%%%%%%%%%%%%%%%%%%%%%%%%%%%%%%%%%%%%%%%%%%%%%%

\newpage
%%%%%%%%%%%%%%%%%%%%%%%%%%%%%%%%%%%%%%%%%%%%%%%%

\newpage
%%%%%%%%%%%%%%%%%%%%%%%%%%%%%%%%%%%%%%%%%%%%%%%%

\newpage
%%%%%%%%%%%%%%%%%%%%%%%%%%%%%%%%%%%%%%%%%%%%%%%%

\newpage
%%%%%%%%%%%%%%%%%%%%%%%%%%%%%%%%%%%%%%%%%%%%%%%%

\end{document}